\documentclass[preprint,12pt,number,sort&compress]{elsarticle}

\usepackage{amsmath}
\usepackage{graphicx}
\usepackage{url}
\usepackage{soul}
\usepackage{caption}
\usepackage{siunitx}
\usepackage{multirow}
\usepackage{enumitem}
\usepackage{float}
\usepackage{lineno}

\newcommand{\parfrac}[2]{\left( \frac{#1}{#2} \right)}
\newcommand{\lami}[1]{\lambda_{#1}}

\journal{arXiv}

\begin{document}

\begin{frontmatter}

\title{Interpreting effective energy barriers to membrane permeation in terms of a heterogeneous energy landscape}

\author[1]{Nathanael S. Schwindt}
\author[2]{Mor Avidar}
\author[2]{Razi Epsztein}
\author[3]{Anthony P. Straub}
\author[1]{Michael R. Shirts\corref{cor1}}
\affiliation[1]{organization={Department of Chemical \& Biological Engineering, University of Colorado Boulder},
                city={Boulder},
                state={CO},
                postcode={80309},
                country={USA}}
\affiliation[2]{organization={Department of Civil and Environmental Engineering, Technion - Israel Institute of Technology},
                city={Haifa},
                postcode={32000},
                country={Israel}}
\affiliation[3]{organization={Department of Civil, Environmental, and Architectural Engineering, University of Colorado Boulder},
                city={Boulder},
                state={CO},
                postcode={80309},
                country={USA}}
\ead{michael.shirts@colorado.edu}
\cortext[cor1]{Corresponding author.}

\begin{abstract} 
    Major efforts in recent years have been directed towards understanding molecular transport in polymeric membranes, in particular reverse osmosis and nanofiltration membranes. Transition-state theory is an increasingly common approach to explore mechanisms of transmembrane permeation with molecular details, but most applications treat all free energy barriers to transport within the membrane as equal. This assumption neglects the inherent structural and chemical heterogeneity in polymeric membranes. In this work, we expand the transition-state theory framework to include distributions of membrane free energy barriers. We show that the highest free energy barriers along the most permeable paths, rather than typical paths, provide the largest contributions to the experimentally-observed effective free energy barrier. We show that even moderate, random heterogeneity in molecular barriers will significantly impact how we interpret the mechanisms of transport through membranes. Simplified interpretations of experimentally measured barriers can lead to incorrect assumptions about the underlying mechanisms governing transport and miss the mechanisms most relevant to the overall permeability. 
\end{abstract}

\begin{graphicalabstract}
\includegraphics[width=\textwidth]{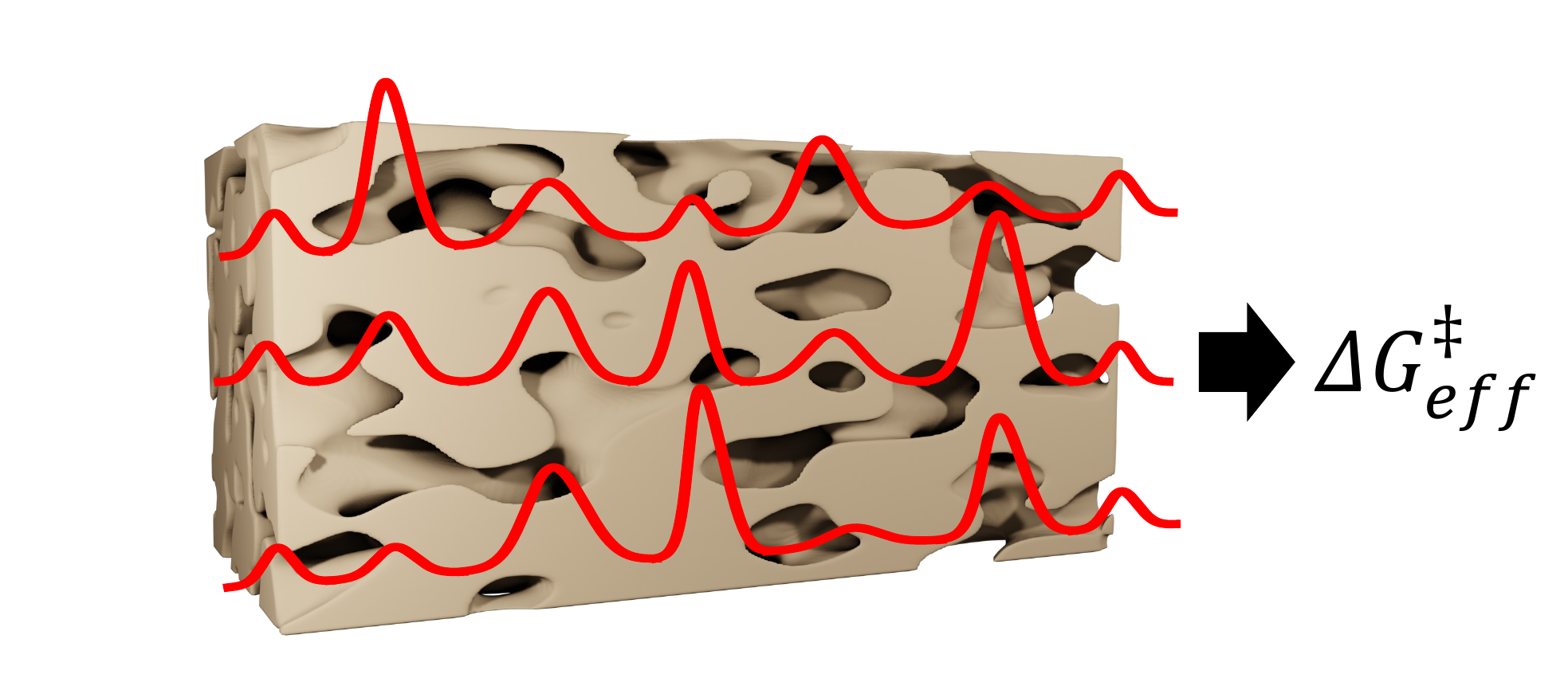}
\end{graphicalabstract}

\begin{highlights}
    \item A novel expression for permeability in terms of heterogeneous free energy barriers is presented.
    \item Measured effective energy barriers are higher than the median barriers in the membrane.
    \item The highest barriers along the most permeable pathways are most representative of measurable energy barriers.
\end{highlights}


\end{frontmatter}


\section{Introduction} \label{s:intro}

Understanding the molecular-level mechanisms that govern transport and selectivity in salt-rejecting membranes, such as those used in nanofiltration (NF) and reverse osmosis (RO), is necessary for the development of next-generation desalination technologies~\cite{epsztein_towards_2020,faucher_critical_2019,epsztein_elucidating_2018}. Numerous models have been proposed over the years to explain the observed transport and selectivity trends in these membranes. However, these models struggle to describe the molecular details of transport through nanometer and sub-nanometer membrane voids and channels~\cite{wang_pore_2021,yaroshchuk_non-steric_2001,biesheuvel_theory_2023,roy_framework_2019}. Developing improved theoretical frameworks and approaches will enhance our understanding of molecular transport in polymeric membranes and help to design future membranes that can address specific requirements~\cite{zhao_differentiating_2021,duchanois_membrane_2021,duchanois_designing_2022}.


As a result, a number of studies of RO and NF membranes have examined the utility of measuring energy barriers to membrane permeability based on either the Arrhenius framework~\cite{zhai_roles_2022,lu_dehydration-enhanced_2023,epsztein_elucidating_2018,zhou_intrapore_2020,epsztein_role_2018} or the similar but more rigorous transition-state theory framework~\cite{shefer_applying_2022,white_theoretical_2022,shefer_enthalpic_2021,rickman_temperature-variation_2014,shefer_limited_2022,kingsbury_kinetic_2024} in order to elucidate details of the molecular mechanisms of transport via experiment. At the simplest level, the Arrhenius activation energy model can be used to understand the energetics of molecular barriers. The Arrhenius equation relates the rate constant $k$ to a pre-exponential factor $A$ and the reaction's activation energy $E_a$, as shown in Eq.~\ref{eq:Arrhenius} where $R$ and $T$ are the gas constant and temperature, respectively.

\begin{equation}
    k = A \exp{\left( \frac{-E_a}{R T} \right)} 
    \label{eq:Arrhenius}
\end{equation}

Based on this framework, the activation parameters (i.e., the energy barrier and the pre-exponential factor) are often measured since they can be directly extracted from the slope and intercept of the linearized Arrhenius equation. By assuming that membrane permeability is a direct function of some molecular-level energy barrier to transport, the Arrhenius equation can be used to estimate these energy barriers. Permeability ($P$), instead of $k$, is treated as an Arrhenius rate in order to relate it to the activation parameters. Linearizing Eq.~\ref{eq:Arrhenius} yields the typical application of the Arrhenius framework for membrane permeability:

\begin{equation}
    \ln P = \ln A -\frac{E_a}{R T}
\end{equation}

These temperature dependent activation parameters could in theory differentiate between mechanisms that are indistinguishable with common modeling frameworks because they are expected to correspond to molecular-level phenomena, such as molecular rearrangement or ion dehydration~\cite{shefer_applying_2022}.

\begin{figure}[ht]
    \centering
    \includegraphics[width=0.75\textwidth]{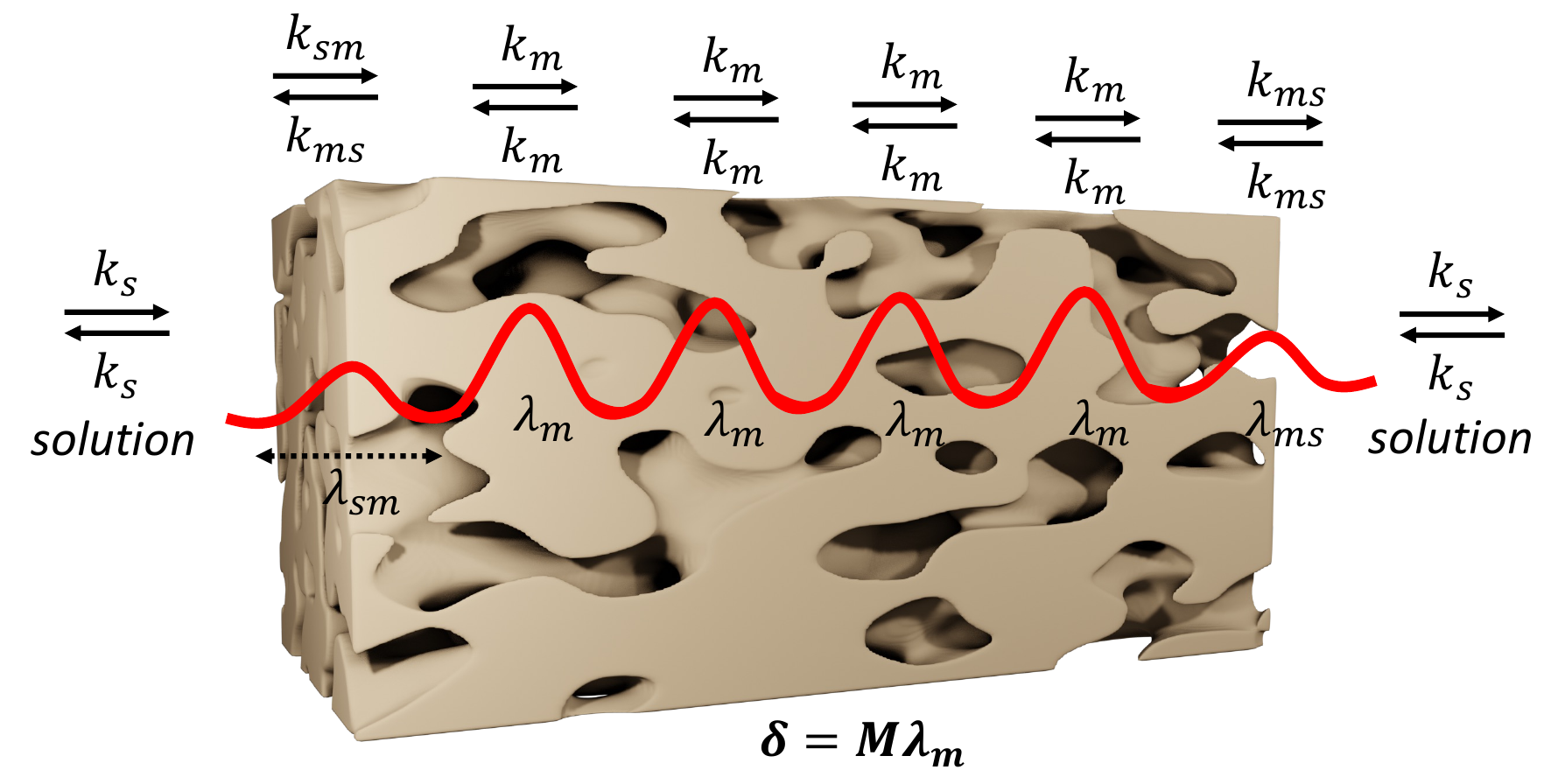}
    \caption{\textbf{Schematic for the membrane model presented by Zwolinski and coworkers~\cite{zwolinski_diffusion_1949}.} $\lambda_{m}$ is the jump length within the membrane, $\lambda_{sm}$ is the jump length for the solution-membrane interface, $\lambda_{ms}$ is the jump length for the membrane-solution interface, $k_s$ is the rate constant for the solution jumps, $k_{sm}$ is the rate constant for the solution-membrane interfacial jump, $k_{ms}$ is the rate constant for the membrane-solution interfacial jump, and $k_m$ is the rate constant for the membrane jump. $M$ is the number of jumps along the transport coordinate, and $\delta$ is the membrane thickness.}
    \label{fig:membrane_illustration}
\end{figure}

A more thermodynamically rigorous model was proposed by Zwolinski, Eyring, and Reese, who described membrane transport using transition-state theory in 1949~\cite{zwolinski_diffusion_1949}, directly connecting permeability to enthalpic and entropic barriers. They adopted Eyring's original theory of reaction rates to describe membrane transport in order to probe how free energy barriers govern permeability. Instead of a quasi-equilibrium between the reactants and the activated complexes, they considered a quasi-equilibrium between molecular jumps through the membrane. They treated membrane transport as jumps governed by rate constants, which could be generalized to any membrane system or transport mechanism, provided that the associated rates were appropriately quantified (Fig.~\ref{fig:membrane_illustration}). They demonstrated the applicability of their framework with biological membranes in a simple solution-membrane-solution framework. 

Typical applications to polymeric membranes rely on the assumption that membrane diffusion can be described as a series of molecular jumps over equal free energy barriers, or equivalently as a single dominant free energy barrier~\cite{del_castillo_energy-barrier_1979,sigurdardottir_energy_2020,zhou_intrapore_2020}. Assuming the barriers within a membrane are equal does not isolate individual mechanisms and ignores the inherent heterogeneity within polymeric membranes. Most notably, such assumption may hinder our understanding of the experimentally measured effective free energy barriers and their associated enthalpic and entropic barriers~\cite{richards_experimental_2013,epsztein_role_2018,shefer_enthalpic_2021}. 

In this study, we extend earlier work on transition-state theory by Eyring and coworkers~\cite{zwolinski_diffusion_1949,giddings_multi-barrier_1958} to account for distributions of free energy barriers that exist within any realistic membrane. Statistical mechanics tells us that to connect molecular phenomena like jumps between voids in a membrane to macroscopic quantities like permeability we must consider the probability distributions associated with those molecular phenomena. We adapt theories for parallel arrays of pores~\cite{wendt_effect_1976,del_castillo_energy-barrier_1979} to molecular pathways through polymeric membranes, developing a novel expression for membrane permeability in terms of molecular jumps along the transport coordinates of many independent pathways. This expression for permeability expands previous derivations~\cite{zwolinski_diffusion_1949,giddings_multi-barrier_1958,scheuplein_application_1968} to generalized membrane barrier distributions. Our mathematical framework is general for any solutes through any membrane, as it is expressed only in terms of transition barriers. However, our choices of parameters, our interpretations, and our conclusions focus on nanofiltration and reverse osmosis membranes for solution-phase separations.

We present a numerical study with statistically random distributions to illustrate the effects of distributions of free energy barriers on the transition-state theory framework. Using this framework, we relate the observable, effective free energy barrier and its enthalpic and entropic components to distributions of energy barriers for individual molecular jumps. An effective free energy barrier that averages molecular events has not been developed previously for arbitrary barrier heights across many parallel paths, despite its growing application in highly heterogeneous polymer membranes. We also explicitly address the accessible area to transport in the derivation of the permeability in terms of the individual molecular barriers across many parallel paths. Finally, we discuss how researchers must use caution when interpreting experimentally observed free energy barriers in membranes, and how heterogeneity even at the molecular level has a significant impact on membrane transport. 

\subsection{Proposed Theoretical Framework}

To construct our framework, we relax two of the main assumptions presented by Zwolinski et al.~\cite{zwolinski_diffusion_1949}, by allowing for distributions of membrane barriers and jump lengths. We apply their equation for flux to a membrane with solution on either side as in Fig.~\ref{fig:membrane_illustration}. We treat all solution jump rates $k_s$ as equal and membrane jump rates $k_{m,j}$ as unequal. Similarly, we treat all solution jump lengths $\lambda_{s}$ as equal and membrane jump lengths $\lambda_{m,j}$ as unequal. As a result, the permeability can be written in terms of the free energy barriers and jump lengths through the interfaces and membrane. The full derivation is provided in the Supplementary Materials Section~\ref{s:SM_permeability_derivation}. We use permeability as it is defined in the original derivation by Zwolinski and coworkers~\cite{zwolinski_diffusion_1949} -- flux divided by concentration gradient. The permeability for a single molecular pathway becomes:

\begin{equation}
    P = \frac{\displaystyle \left( \frac{\lambda_{sm}}{\lambda_{ms}} \right) \left( \frac{k_B T}{h} \right) \exp{\left( \frac{ -\left( \Delta G_{sm}^{\ddagger} - \Delta G_{ms}^{\ddagger} \right)}{R T} \right)}}{\displaystyle \sum_{j=1}^M \left( \frac{1}{\lambda_{m,j}} \right) \exp{\left( \frac{-\Delta G_{m,j}^{\ddagger}}{R T} \right)}}
    \label{eq:permeability_barriers}
\end{equation}

\noindent where $P$ is permeability, and $k_B$, $T$, $h$, and $R$ are Boltzmann's constant, temperature, Planck's constant, and the gas constant. $\lambda_{sm}$ and $\lambda_{ms}$ are the jump lengths from solution to membrane and membrane to solution, respectively. Similarly, $\Delta G_{sm}^{\ddagger}$ and $\Delta G_{ms}^{\ddagger}$ are the free energy barriers for the solution-to-membrane jumps and membrane-to-solution jumps. $\Delta G_{m,j}^{\ddagger}$ is the free energy barrier for membrane jump $j$.

The permeability in Eq.~\ref{eq:permeability_barriers} only describes transport along a single molecular pathway. The observed permeability is a combination of all accessible molecular paths, similar to the parallel array of pores described by Wendt et al.~\cite{wendt_effect_1976}. We apply this relationship to our expression for permeability to obtain an area-weighted permeability across many parallel paths. By introducing the fraction of accessible area, the transition-state theory framework can be applied to both membranes with permanent pores or with fluctuating voids. Therefore, the overall permeability for $N$ paths per unit area each with $M_i$ barriers is: 

\begin{equation}
    P = \sum_{i=1}^{N} \left[ \frac{\displaystyle \left( \frac{A_i}{A_0} \right) \left( \frac{\lambda_{sm}}{\lambda_{ms}} \right) \left( \frac{k_B T}{h} \right) \exp{\left( \frac{ -\left( \Delta G_{sm}^{\ddagger} - \Delta G_{ms}^{\ddagger} \right)}{R T} \right)}}{\displaystyle \sum_{j=1}^{M_i} \left( \frac{1}{\lambda_{m,i,j}} \right) \exp{\left( \frac{-\Delta G_{m,i,j}^{\ddagger}}{R T} \right) }} \right]
    \label{eq:overall_permeability}
\end{equation}

\noindent where $A_i$ is the cross-sectional area for path $i$, $A_0$ is the membrane unit area being considered, and $M_i$ is the number of membrane jumps for path $i$. Similar to Eq.~\ref{eq:permeability_barriers}, $\lambda_{m,i,j}$ and $\Delta G_{m,i,j}^{\ddagger}$ are the jump length and free energy barrier for the $j^{th}$ membrane jump on path $i$, respectively.

We express the effective free energy barrier from Eyring's original derivation (Eq.~S5) in terms of distributions of membrane free energy barriers and jump lengths across many parallel paths with different numbers of jumps. To do this, we equate Eq.~S5 to Eq.~\ref{eq:overall_permeability} and solve for $\Delta G_{eff}^{\ddagger}$. Eq.~\ref{eq:effective_barrier} gives the resulting analytical expression for the overall effective free energy barrier, the main theoretical result of this paper.

\begin{equation}
    \Delta G_{eff}^{\ddagger} = -R T \ln{\left[ \sum_{i=1}^{N} \frac{ \displaystyle \left( \frac{A_i}{A_0} \right) \left( \frac{\delta}{\lambda_{avg}^2} \right) \left( \frac{\lambda_{sm}}{\lambda_{ms}} \right) }{\displaystyle \sum_{j=1}^{M_i} \left( \frac{1}{\lambda_{m,i,j}} \right) \exp{\left( \frac{\Delta G_{m,i,j}^{\ddagger}}{R T} \right) }} \right]} + \left( \Delta G_{sm}^{\ddagger} - \Delta G_{ms}^{\ddagger}\right)
    \label{eq:effective_barrier}
\end{equation}

We can decompose this effective free energy barrier into enthalpic and entropic terms. Under the same assumptions as the original expression by Zwolinski et al.~but expanded to include parallel paths, we find an effective entropic penalty resulting from the fraction of membrane area accessible to permeation. The permeability only depends on the path areas that are accessible to transport. The accessible area to transport is not necessarily the total membrane area, as shown in Eq.~\ref{eq:SM_parallel_permeability}. This result is consistent with experimental barriers calculated for ions in NF membranes, where the entropy was attributed to geometric constraints on the void volumes~\cite{epsztein_elucidating_2018,shefer_enthalpic_2021}. Eq.~\ref{eq:entropic_penalty} more clearly shows this ``entropic'' penalty if we additionally assume all paths are identical. Zwolinski et al.~implicitly assumed that the entire area is accessible to transport, or equivalently that $\sum_{i=1}^{N} A_i = A_0$ such that the entropic penalty is 0. The area fraction accessible to transport, because it is not temperature dependent, would manifest as part of the overall effective entropy. When we expand the scenario presented by Zwolinski et al.~to parallel paths, the overall equation becomes:
\begin{equation}
    \Delta G_{eff}^{\ddagger} = \left[ \Delta H_{m}^{\ddagger} + \Delta H_{sm}^{\ddagger} - \Delta H_{ms}^{\ddagger} \right] - T \left[ \Delta S_{m}^{\ddagger} + \Delta S_{sm}^{\ddagger} - \Delta S_{ms}^{\ddagger} + R \ln{ \left( \sum_{i=1}^N \frac{A_i}{A_0} \right) } \right] 
    \label{eq:entropic_penalty}
\end{equation}

\noindent $\Delta H_m^{\ddagger}$ and $\Delta S_m^{\ddagger}$ are the enthalpic and entropic barriers within the membrane, and similarly, $\Delta H_{sm}^{\ddagger}$, $\Delta S_{sm}^{\ddagger}$, $\Delta H_{ms}^{\ddagger}$, and $\Delta S_{ms}^{\ddagger}$ are the enthalpic and entropic barriers at the solution-membrane ($sm$) and membrane-solution ($ms$) interfaces.

In this study, we focus on the scenario where transport is primarily hindered by diffusion through the membrane, not membrane entry or exit~\cite{zhou_intrapore_2020,song_molecular_2020}. As a result, we treat the jumps across the solution-membrane and membrane-solution interfaces as fast and their associated free energy barriers as negligible. We additionally assume interfacial barriers are constant across all parallel paths. In this case, $\Delta G_{sm}^{\ddagger}$ and $\Delta G_{ms}^{\ddagger}$ are constant for all $i$ paths and small compared to $\Delta G_{m,i,j}^{\ddagger}$. However, in some cases, these barriers may be significant factors in modeling membrane transport. For example, ion transport through charged membranes may introduce a large barrier due to Donnan exclusion~\cite{kingsbury_kinetic_2024}. To include the effect of the interfaces, there are two significant scenarios to address. In the first case, interfacial barriers dominate the transport. Only the interfacial barriers and their heterogeneity across parallel paths would need to be considered. In the second case, interfacial barriers are of similar magnitude to barriers within the membrane. Since the individual barriers appear as a sum in the permeability expression, the order of barriers does not change the interpretation of the single path effective free energy barrier~\cite{scheuplein_application_1968}. Therefore, the interfacial barriers can be included in the overall framework with corresponding distributions.

\section{Experimental}

\subsection{Numerical methods} \label{s:numerical_methods}

We numerically evaluate our expanded transition-state theory model for membrane permeability by drawing magnitudes for each of the individual enthalpic and entropic barriers from independent random distributions. To explore a range of resulting outcomes, we select two common distributions with some physical motivation. First, we assume a fixed mean and normally distributed barrier heights around this mean. Physically, this distribution would model membranes with a consistent nanostructure on average, with some statistical variation at the molecular level. Most molecular pathways would thus have similar environments and jump mechanisms, such that the barriers would be similar, though with some variation. Second, we choose exponentially distributed barrier heights to represent membranes with a large amount of heterogeneity. All paths would have regions of unfavorable mechanisms with a few high barriers, as well as regions of low-barrier mechanisms more similar to free diffusion. Normal and exponential distributions occur in many natural phenomena, and thus represent two useful extremes of possible behavior. In virtually all cases, the actual distributions best describing membrane transport are unknown, and we may not have explored the parameter ranges that are most physically relevant. However, the analysis is generalized to be broadly applicable, so the main conclusions of the work are unaffected by the choice of distributions. 

The free energy, enthalpy, and entropy associated with a molecular jump are interrelated; only two can be specified independently. We draw enthalpic and entropic barriers from independent distributions. In reality, these barriers are likely correlated, for example, through observed enthalpy-entropy compensation. However, it would be difficult to estimate appropriate covariances as enthalpy-entropy compensation is not well-understood in polymeric membranes~\cite{shefer_limited_2022, kingsbury_kinetic_2024}. We draw heights of the enthalpic barriers from distributions with mean 3.5 kcal/mol, which corresponds to the observed effective enthalpic barrier for chloride within NF membranes at 300~K~\cite{pavluchkov_indications_2022}, and we draw entropic barriers from distributions with mean -0.03 kcal/mol$\cdot$K, which corresponds to the observed effective entropic barrier for chloride under the same conditions~\cite{pavluchkov_indications_2022}. This combination results in an effective free energy of $\Delta G_{eff}^{\ddag} = \Delta H_{eff}^{\ddag} - T \Delta S_{eff}^{\ddag}$ = 12.5 kcal/mol, at 300~K. Unless otherwise specified, the standard deviation for the normally distributed enthalpic barriers is 1.17 kcal/mol, and the standard deviation for the normally distributed entropic barriers is 0.01 kcal/mol$\cdot$K. These standard deviations ensure the normally distributed barriers represent membranes with less heterogeneity than the exponentially distributed barriers. Exponential distributions are defined by a single parameter, so specifying their mean is enough to fully define them.

Typical RO and NF membrane selective layers are between 10 and 200~nm, or 100 and 2000~\AA ~\cite{chowdhury_3d_2018, cadotte_new_1980}. We estimate individual jumps to be between 1 and 10~\AA~as done in previous work based on diffusion calculations~\cite{shefer_applying_2022, zwolinski_diffusion_1949, rickman_temperature-variation_2014}. Assuming no tortuosity along the path results in 10 to 2000 jumps. We use 200 jumps of length 2~\AA~unless otherwise specified. We test how sensitive our results are to jump lengths and number of jumps in Supplemental Materials Section~\ref{s:jump_distributions}. 

In simulating membranes with multiple paths across the membrane, we use \num{2e-4} as an estimate for the number of paths per $\text{\AA}^2$. This estimate is approximately one order of magnitude smaller than the estimated packing density of single-walled carbon nanotubes with a diameter of 0.5 nm (more information provided in the Supplementary Materials Section~\ref{s:estimate_paths}), to account for the heterogeneity of polymer membranes. Here, we show trends for 2000 paths through the membrane unless otherwise stated, which roughly corresponds to a unit area of 0.1 \unit{\um\squared}, enough to converge average results across a distribution of paths. See Fig.~\ref{fig:convergence} for determination of the number of paths needed for convergence. When testing the model, we assume independent and separate pathways through the membrane, but in reality, the molecular-level pathways almost certainly can merge, split, and interconnect. Incorporating this additional heterogeneity is beyond the scope of the current study. However, this framework can be easily extended to introduce correlated barrier distributions between paths and consideration of topological effects, as for example in the work of Culp et al.~\cite{culp_nanoscale_2021}. Similar to correlations between barriers, it would be difficult to determine \textit{a priori} appropriate covariances for any given polymeric membrane. However, by treating the molecular jumps as resistances, many different topologies could be explored with parallel-series circuit models. This circuit model theory is well-developed for interconnected pathway flows and can be readily expanded to include varying barriers ~\cite{mishra_effective_2021,zhang_equivalent_2019,tan_resistance_2017,tan_equivalent_2013}. The code implementation for our numerical analysis is on Github at \url{https://github.com/shirtsgroup/eyring_model}. 

\subsection{Crossflow filtration experiments}

Filtration experiments were performed with two types of flat-sheet commercial membranes in a crossflow mode -- a loose polyamide NF membrane (NF270, Dow FilmTec) and a tight polyamide RO membrane (SW30, Dow FilmTec). Single-salt solutions of NaCl and NaF at 5~mM were used as feed solutions. The filtration experiments were carried out at pH~7, with an applied pressure of 33~bar and a crossflow velocity of 2.13~m/s. In order to calculate transition-state theory barriers, the salt flux was measured at 6 temperatures from 10 \textdegree C to 40 \textdegree C. The permeability at these temperatures was calculated using
\begin{equation}
    P = \frac{J_s}{C_m - C_p}
    \label{eq:experimental_permeability}
\end{equation}
\noindent where $J_s$ is the salt flux and $C_m$ and $C_p$ are the salt concentrations on the membrane surface in the feed side and in the permeate solution, respectively. Concentration polarization on the membrane surface and $C_m$ were evaluated using previously reported methods given also in the Supplementary Materials Section~\ref{s:concentration_polarization}~\cite{peer-haim_adverse_2023}. The effective overall enthalpic and entropic barriers were extracted from the slope and intercept of the linearized Eyring plot, as shown previously~\cite{shefer_limited_2022}. These experiments were replicated 3 times for each temperature.

\subsection{Measurement of barriers to transport for salt in water}

Energy barriers to transport of salt in water were calculated by measuring the conductivity (Eutech Instruments, CON2700) of 5~mM sodium chloride solutions at 4 different temperatures between 25 \textdegree C and 45 \textdegree C. The barriers for the conductivity were then extracted using the same method applied to calculate the barriers of the permeability. The resulting transition-state theory plot for conductivity in water is provided in the Supplementary Materials Fig.~\ref{fig:tst_NaCl_water}.

\section{Results}

\subsection{The single path effective free energy barrier is highly dependent on the heterogeneity of the individual barrier distributions}

\begin{figure}[ht!]
    \centering
    \includegraphics[width=\textwidth]{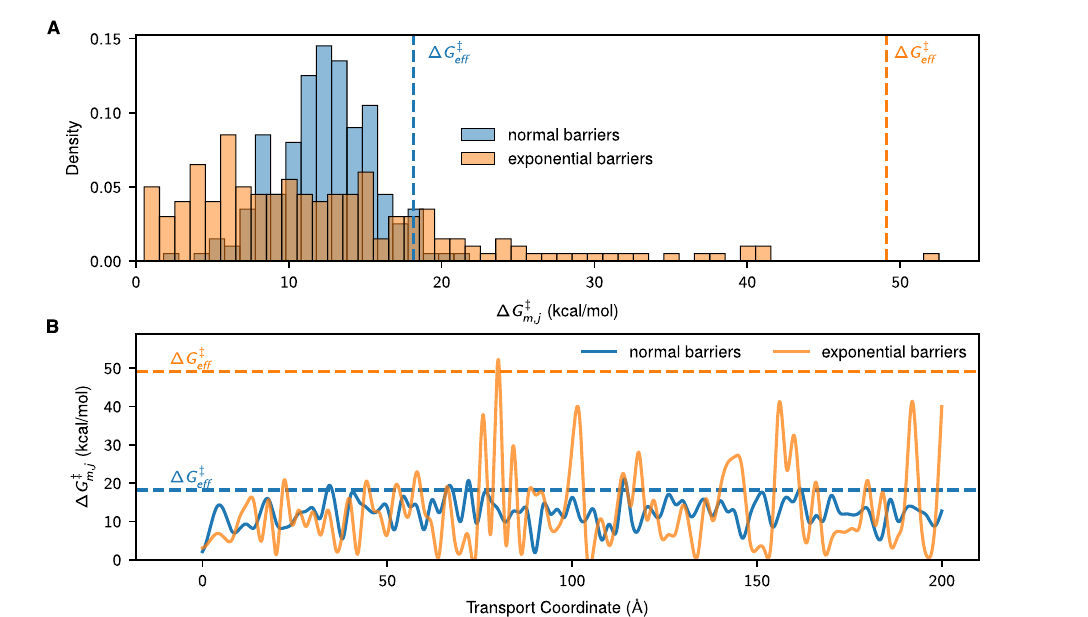}
    \caption{\textbf{A realization of distributions of membrane barriers along a single path.} (\textbf{A}) For both realizations considered, the effective free energy barrier for a single path lies near the maximum of the distribution. Larger variance in the distribution results in a significantly larger effective barrier. The effective free energy barrier is shown as a dashed vertical line. The mean free energy barrier for both distributions is 12.5 kcal/mol with further discussion in Section~\ref{s:numerical_methods}. We use 200 jumps of 2 \AA~each through a single path at 300 K. (\textbf{B}) The effective free energy barrier along a single path is most similar to the maximum barrier along the path. We show only half of the membrane pathways simulated in \textbf{A} to better visualize the individual barriers. The effective free energy barriers for each path are shown as dashed horizontal lines. Enthalpic and entropic barriers are each drawn independently from the specified distributions and combined to calculate the free energy barrier.}
    \label{fig:effective_barriers}
\end{figure}

We find that the effective free energy barrier along a single path is slightly below the maximum free energy barrier of the underlying distributions, and significantly larger than the mean free energy barrier. The single path effective free energy barrier can be related to the distribution of membrane free energy barriers $\Delta G_{m,j}^{\ddagger}$ by assuming a single path $i$ where the entire area is accessible to transport. Fig.~\ref{fig:effective_barriers}A shows where the single path effective free energy barrier would lie for one realization of the barrier distribution, assuming the underlying distributions for the enthalpic and entropic barriers follow normal distributions and exponential distributions. Fig.~\ref{fig:effective_barriers}B shows free energy profiles for single pathways through the membrane with the barrier distributions in Fig.~\ref{fig:effective_barriers}A. We only show free energy profiles for half of the length of the membrane to ensure the figure is legible. 

The single path effective free energy barrier is most affected by the largest individual barriers, qualitatively consistent with Giddings and Eyring's kT-cutoff~\cite{giddings_multi-barrier_1958}. We numerically test the kT-cutoff model by comparing the effective free energy barrier calculated with all barriers and the effective free energy barrier calculated with only those in the kT-cutoff. The effective barrier calculated with the barriers within the kT-cutoff is within 15\% and 6\% of the actual effective barrier, for 1000 realizations of normally distributed and exponentially distributed barriers, respectively. Larger variance in the underlying barrier distributions introduces high outliers that significantly increase the single path effective free energy barrier. In Fig.~\ref{fig:effective_barriers}A, the higher variance path with exponentially distributed barriers gives a much larger single path effective barrier than the path with normally distributed barriers. Fig.~\ref{fig:effective_barriers} shows that the single path effective barrier is slightly below the maximum barrier and well above the mean at 12.5 kcal/mol. The effective barrier through a single path does not depend on the locations or orders of the barriers, as it can be calculated from an unordered distribution as in Fig.~\ref{fig:effective_barriers}A.

\subsection{The overall effective free energy barrier is determined by the highest barriers in the most permeable paths}

\begin{figure}[ht!]
    \includegraphics[width=\textwidth]{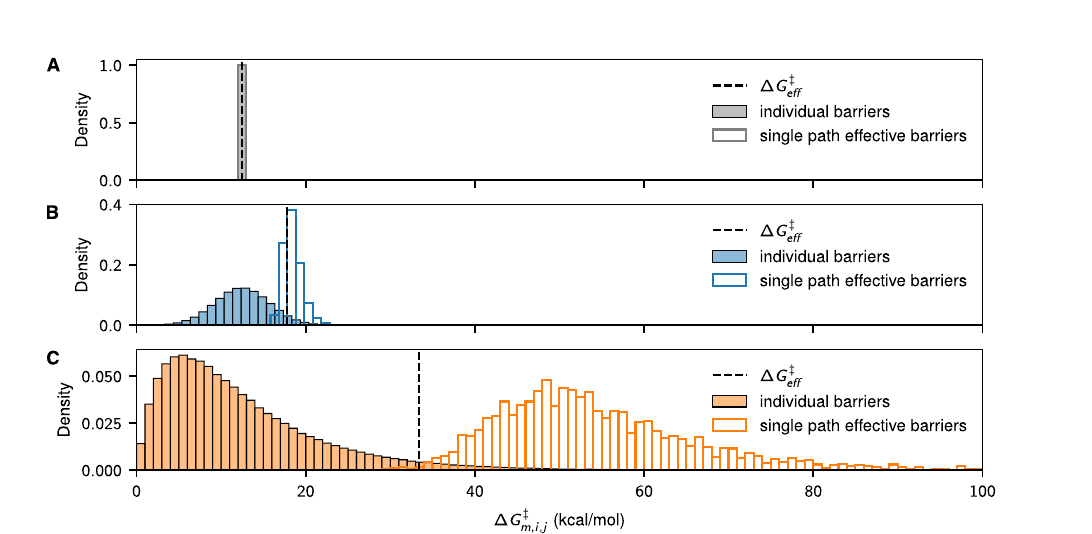}
    \caption{\textbf{Effective free energy barriers depend on the largest underlying barriers.} The overall effective free energy barrier is near the maximum of the individual free energy barriers and near the minimum of the single path effective free energy barriers. Membrane barrier distributions have  the same mean of 12.5 kcal/mol. The underlying enthalpic and entropic barriers are all equal (\textbf{A}), normally distributed (\textbf{B}), and exponentially distributed (\textbf{C}). The effective free energy barriers are shown as dashed vertical lines. The overall effective free energy barriers are calculated by Eq.~\ref{eq:effective_barrier}. We use 200 jumps of 2 \AA~each for all 2000 paths.}
    \label{fig:effective_barriers_parallel}
\end{figure}

Expanding the model to a membrane comprising many parallel paths, we find the overall effective free energy barrier through the membrane from Eq.~\ref{eq:effective_barrier} lies within the high tail of the underlying barrier distributions and the low tail of the single path effective barriers. In Fig.~\ref{fig:effective_barriers_parallel}, we show the overall effective free energy barriers for 2000 paths compared to the distributions of individual free energy barriers and the distributions of single path effective barriers. The entropic penalty from the accessible area for transport as shown in Eq.~\ref{eq:entropic_penalty} is assumed to be 0. If all paths have equal individual barriers as in the original Zwolinski et al.~derivation, the overall effective free energy barrier collapses to be identical to an individual membrane barrier, as shown in Fig.~\ref{fig:effective_barriers_parallel}A. 

\begin{figure}[ht!]
    \centering
    \includegraphics[width=\textwidth]{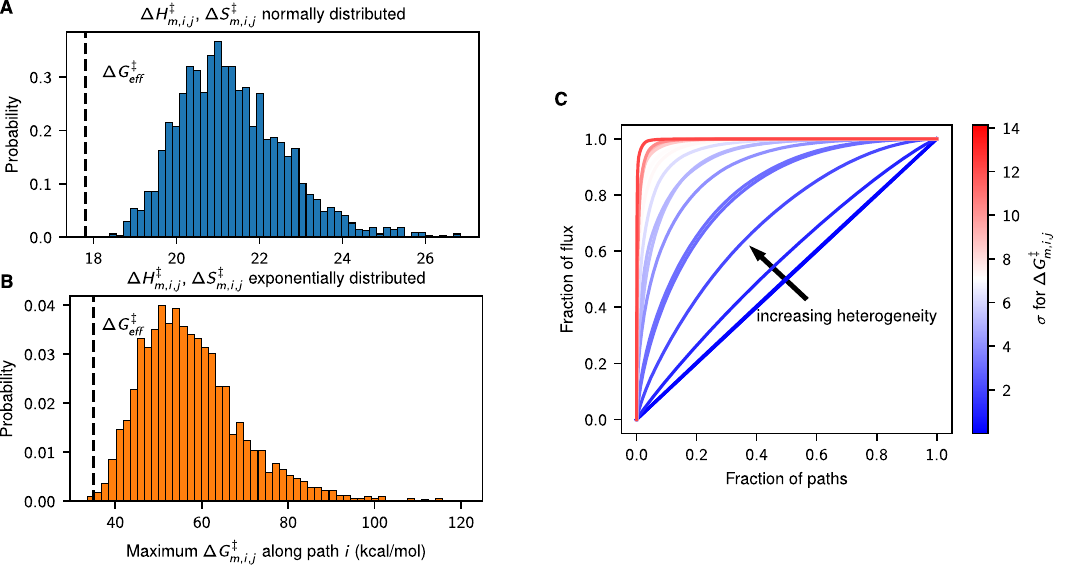}
    \caption{\textbf{The effective free energy barrier and flux are dominated by paths with low maximum barriers.} (\textbf{A},\textbf{B}) The effective free energy barrier in the case of many pathways is primarily determined by the paths with the smallest maximum barriers. We show the distribution of maximum barriers for each of 2000 paths through the membrane. For normally (\textbf{A}) and exponentially (\textbf{B}) distributed enthalpic and entropic barriers, the overall effective free energy barrier is shown as a dashed line. The entropic penalty from the accessible area for transport is assumed to be 0 to highlight the effect of the barrier heights. (\textbf{C}) More heterogeneous free energy landscapes create a few highly permeable paths that dominate the flux, as shown by the fraction of the flux through the most permeable paths. The fluxes are calculated in the case of normally distributed enthalpic and entropic barriers with increasing variance as given by the standard deviation ($\sigma$) of $\Delta G_{m,i,j}^{\ddagger}$. The standard deviation for the enthalpic barriers ranges from \num{1e-4} to 10 kcal/mol, and the standard deviation for the entropic barriers ranges from \num{3.3e-7} to \num{3.3e-2} kcal/mol$\cdot$K.}
    \label{fig:max_barriers_and_ROC}
\end{figure}

Fig.~\ref{fig:effective_barriers_parallel}B and Fig.~\ref{fig:effective_barriers_parallel}C show that the overall effective barrier for the membrane lies near the maximum individual barrier. Equivalently, the overall effective barrier lies near the lowest single path effective barrier. Therefore, when we consider distributions of free energy barriers across many pathways, the overall effective free energy barrier to permeability is not the difference in free energy between the species in solution and the species at the top of the highest potential energy barrier. Rather, it is heavily dependent on the highest barriers within the paths with the lowest single path effective barriers. These single path effective barriers are most dependent on the highest individual barriers along the path. We demonstrate that the overall effective barrier is typically determined by the paths whose highest barriers are relatively low in Fig.~\ref{fig:max_barriers_and_ROC}A and Fig.~\ref{fig:max_barriers_and_ROC}B, where we plot each path's maximum barrier. The overall effective barrier is near the lowest maximum barriers, which is in turn near, but not at, the top of the distribution of individual barrier heights.

\subsection{Heterogeneity in molecular pathways dictates membrane flux}

Intuitively, the overall flux is most determined by the paths with the highest permeability, and Fig.~\ref{fig:max_barriers_and_ROC}C confirms this trend in the transition-state theory model for molecular pathways through a membrane. Importantly, this is true not only for macroscopic defects, but also for mechanistic molecular barriers. If all individual membrane barriers are equal, the flux is evenly distributed across all parallel paths as shown in the straight, dark blue line. As more heterogeneity is introduced from the distributions of membrane barriers, the flux is skewed towards highly permeable paths. Distributions of free energy barriers within the membrane create more favorable paths through the membrane. Physically, paths through easily traversed voids will contribute most to the total permeability, and paths that require energetically unfavorable rearrangement and hopping will contribute least to the permeability. 

\begin{figure}[ht!]
    \includegraphics[width=0.9\textwidth]{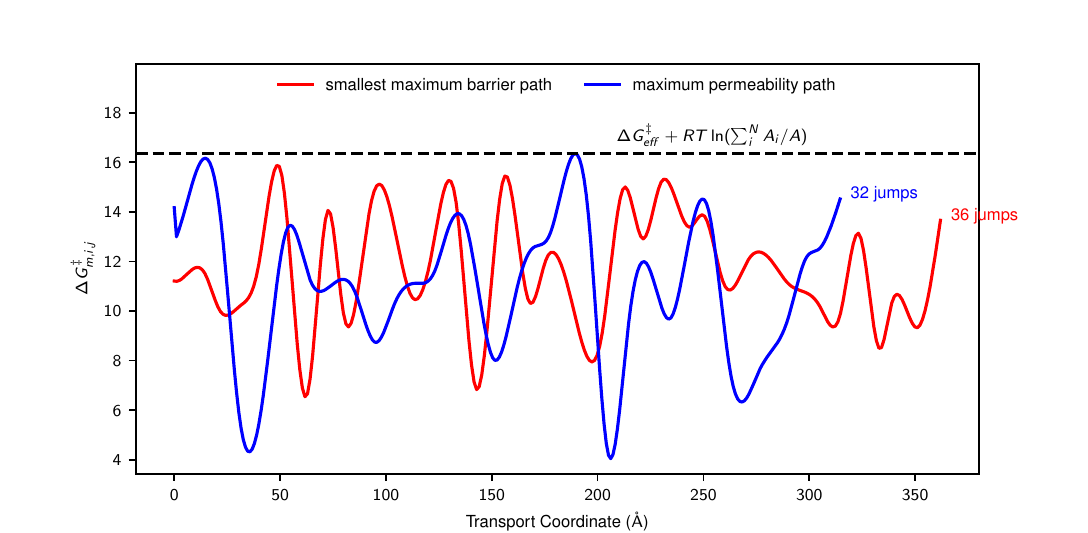}
    \caption{\textbf{Smallest maximum barrier path and the most permeable path through a membrane with normal barriers, jump lengths, and jump numbers.} Pathways with low maximum barriers and a few large jumps contribute most to the overall effective barrier. Over a realization of 2000 paths with normally distributed enthalpic and entropic barriers, jump lengths, and number of jumps, the path with the smallest maximum barrier is shown in red, and the path with the highest permeability is shown in blue. While the most permeable path does not have the smallest maximum barrier, its maximum barrier is low and it requires fewer jumps. The overall effective free energy barrier for 2000 paths, shifted by the effective entropic penalty from parallel paths is shown as a dashed line. We shift the effective free energy barrier by the entropic penalty to better show the direct connection of effective barriers to the barrier height distribution.}
    \label{fig:vary_everything}
\end{figure}

For a real membrane, all the molecular pathways will have variance in their energy barriers, jumps, and total path length, and therefore outlier pathways with high permeability will contribute the most to observable energy barriers. Fig.~\ref{fig:vary_everything} gives one realization of the model where the underlying barriers, jump lengths, and the number of jumps are each normally distributed, and we highlight two important free energy profiles through the membrane---the most permeable path (blue) and the path with the smallest maximum barrier (red). The highest permeability paths have low maximum barriers and fewer jumps. In the model, fewer jumps corresponds to fewer opportunities for high outliers in the membrane barrier distribution. These paths might correspond to large voids or defects in the membrane, where molecules can easily take large jumps. A low maximum barrier may represent a pore that has a single small constriction but is otherwise relatively open. Another high permeability path may be through a region of loose, flexible polymer, where most polymer rearrangements are low-energy or allow for large jumps.

In contrast to the substantial effect from distributions in the barrier thermodynamics, we find the overall effective free energy barrier does not vary much with distributions of number of jumps or jump lengths. Assuming all barrier heights are equal, introducing heterogeneity in the jumps results in changes on the order of 0.5 kcal/mol. This finding has a caveat that the overall effective barrier can be decreased moderately when a non-negligible number of paths contain only a few (single digit numbers) jumps, where the effects might be as large as 1.5 kcal/mol. However, this is still a small contributing factor compared to the effects from variation in barrier heights. In the case of varying \textit{both} barrier heights and numbers of jumps, we expect a larger difference in the overall effective barrier caused by paths with both small numbers of jumps and no high energy barriers among those jumps. We do not explore this regime quantitatively in this study due to the large number of possible variables. We present an in-depth discussion of the effects of jump lengths and number of jumps in the Supplementary Materials Section~\ref{s:jump_distributions}.

In Fig.~\ref{fig:vary_everything}, we show a realization where the most permeable path does not have the smallest maximum barrier. While its maximum barrier is comparatively small, it is not the smallest maximum barrier. We tested how frequently the smallest maximum barrier path is also the most permeable path for both normally distributed and exponentially distributed underlying barriers. For 1000 realizations of normally distributed membrane barriers, the smallest maximum barrier path is the most permeable path in 60.8\% of the realizations. That percentage jumps to 90.0\% for exponentially distributed barriers with the same mean. Of the realizations where the smallest maximum barrier path is not the most permeable path, the maximum barrier in the most permeable path is similar to the smallest maximum barrier 95.9\% of the time for normally distributed barriers and 97.0\% of the time for exponentially distributed barriers. Barriers are considered similar if they are within $k_B T$, as defined by Giddings and Eyring's kT-cutoff~\cite{giddings_multi-barrier_1958}.

\subsection{Overall effective enthalpic and entropic barriers are larger than the typical barriers experienced by molecules in the membrane}

\begin{figure}[ht!]
    \includegraphics[width=0.9\textwidth]{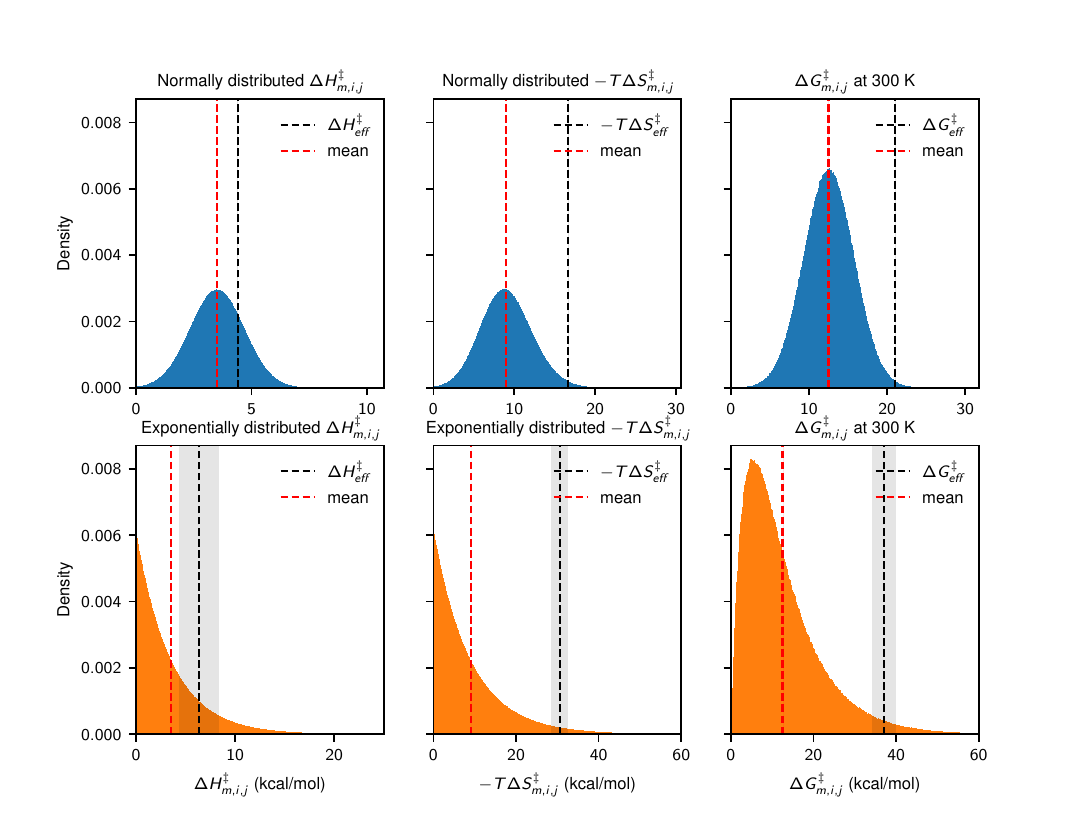}
    \caption{\textbf{Distributions of enthalpic, entropic, and free energy barriers and their overall effective barriers.} Effective entropic and enthalpic barriers are found towards the tails of their underlying distributions. In these examples, the effective entropic barriers have a larger relative shift compared to the effective enthalpic barriers due to the accessible area entropic penalty. We calculate the overall effective barriers for 22,000 paths from the linearized permeability vs. temperature. Standard errors for the effective barriers and means are shown as lightly shaded regions. Error for the means and normally distributed barriers are too small to be visible.}
    \label{fig:dH_dS_distributions}
\end{figure}

Zwolinski and coworkers' expression for permeability (given in our Eq.~S5) has been used to estimate the overall effective enthalpic and entropic barriers to membrane transport. Typically, this equation is linearized so the slope is $-\frac{\Delta H_{eff}^{\ddagger}}{R}$ and the intercept is $\frac{\Delta S_{eff}^{\ddagger}}{R}$. Therefore, the enthalpic and entropic contributions to the permeability can be estimated by simply measuring permeability at a range of temperatures~\cite{shefer_applying_2022}. We follow this approach using permeabilities from our numerical model evaluated at a range of temperatures. Individual barriers at each temperature are drawn from random distributions with the same parameters.

As with free energies, the overall effective enthalpy and entropy calculated from the linearized fit lie in the high magnitude tail of the underlying distributions of enthalpies and entropies. Fig.~\ref{fig:dH_dS_distributions} demonstrates how the measured enthalpic and entropic barriers are larger in magnitude than their respective average barriers in the membrane. These distributions are for 22,000 paths with 200 jumps each for temperatures at 10~K increments between 250~K and 350~K. We simulate 22,000 paths (which would be approximately equivalent to 1 \unit{\micro\meter\squared}, following the same procedure provided in the Supplementary Materials Section~\ref{s:estimate_paths}) to reduce the error in the effective barriers for the exponential distributions, since the exponential distributions have higher variance. 

The relative shift in the entropic barrier is larger than that for the enthalpic barrier because the entropic shift also includes the contribution from transport-accessible area, as that contribution is temperature independent and would be interpreted as entropy. Additionally, in Fig.~\ref{fig:dH_dS_distributions}, the mean entropic contribution $\left< -T\Delta S_{m,i,j}^{\ddagger} \right>$ is larger than the enthalpic contribution, further exaggerating the difference in effective barriers. Increasing the variance in the membrane barrier distribution increases the magnitudes of the overall effective enthalpic and entropic barriers, as shown by the higher variance exponential distributions in Fig.~\ref{fig:dH_dS_distributions}. Higher variance introduces higher maximum barriers along single paths, which heavily influences the overall barrier to transport. We find that observable entropic and enthalpic barriers are, again, not representative of the typical or mean mechanisms in the membrane but rather, of the rate-limiting mechanisms along only the most permeable paths.

\subsection{Implications for experimental study of effective energy barriers in RO and NF membranes}

\begin{figure}[ht]
    \centering
    \includegraphics[width=\linewidth]{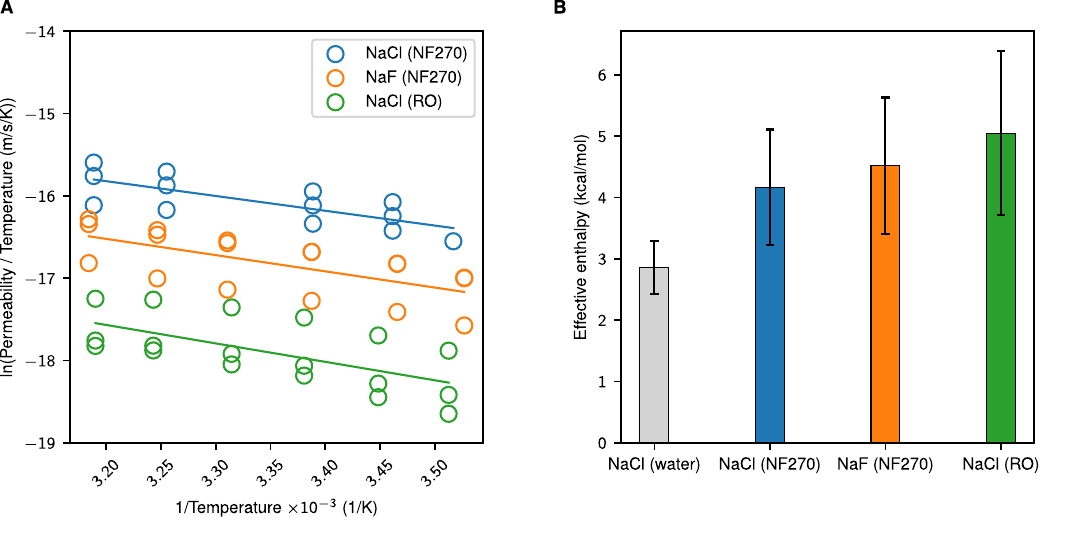}
    \caption{\textbf{Experimental linearized transition-state theory plots and the resulting overall effective enthalpic barriers.} Overall effective barriers measured experimentally are similar across different salts and membranes, indicating that the highest barriers in the most permeable paths are also similar, despite changes in membrane and salts. (\textbf{A}) Linearized transition-state theory plots for the permeability of NaCl and NaF in the NF270 membrane and NaCl in the SW30 RO membrane. The least squares fit is shown as a line for each system. (\textbf{B}) Overall effective enthalpic barriers calculated from the slopes in (\textbf{A}) and Fig.~\ref{fig:tst_NaCl_water}. The errors shown are the propagated errors from the linear regression. Experimental conditions during filtration: a single-salt solution of NaCl or NaF at 5 mM, 10-40 \textdegree C, pH 7, 33 bar, and crossflow velocity of 2.13 m/s. }
    \label{fig:experimental_data}
\end{figure}

In experimental studies of molecular transport in polymeric membranes, the measured barrier is considered an overall effective parameter that represents the transport of a given solute, and the physical meaning has not been fully established for aqueous transport in polymeric membranes. Based on the current study, we can better analyze and understand effective energy barriers in the context of many individual energy barriers in parallel and series.

To demonstrate the implications our analysis has on the experimental study of effective energy barriers, we extracted effective transition state barriers from permeabilities of a selection of salts in a selection of membranes.  Specifically, 
we experimentally measured the permeability of sodium chloride (NaCl) in a NF membrane at six temperatures to extract the effective enthalpic barrier for the salt transport from the slope of the linearized transition-state theory plot (Fig.~\ref{fig:experimental_data}A). We also performed a similar measurement for the transport of sodium fluoride (NaF) in the same NF membrane and for NaCl in a RO membrane (Fig.~\ref{fig:experimental_data}A). 
Finally, we measured the increase of NaCl conductivity with temperature in water and constructed its corresponding linearized transition-state theory plot (Fig.~\ref{fig:tst_NaCl_water}). The effective enthalpic barriers measured for the four cases are shown in Fig.~\ref{fig:experimental_data}B.

Fig.~\ref{fig:experimental_data}B does show an increasing effective enthalpic barrier to transport with a denser medium (water $<$ NF $<$ RO) or a larger and more strongly hydrated species (NaCl $<$ NaF). These trends are intuitive as a denser medium or larger species may require higher molecular adjustments and arrangements during diffusion jumps. The effective enthalpic barriers measured for membrane permeability are slightly higher than typical barrier values reported for water or ion diffusion in water,~\cite{wang_self-diffusion_1953,talekar_temperature_2009} indicating a hindered diffusion compared to free diffusion in water. However, the differences in the enthalpic barrier heights are within the statistical uncertainty, so the effective enthalpies do not appear to be significantly affected by the substantial change in ion size, nor by membrane density. This observation supports the picture that the average ion environment only loosely affects the transport along the most important paths. For example, both the NF and RO membranes may have low density paths or large, interconnected voids that dominate the flux, resulting in similar effective barriers, despite the significant difference in the chemistry of the membranes.


Our finding that the overall effective energy barrier is dictated by the highest barrier in the most permeable path is also supported by prior experimental data examining the heterogeneity of polyamide RO membranes~\cite{culp_nanoscale_2021}.  Culp et al.~identified water diffusion pathways in polyamide RO membranes and estimated the local flux along those pathways. They found that the average diffusion coefficient in the polymer was unable to predict membrane water permeability, rather that the nanoscale heterogeneity controlled membrane permeability. They identified the same two levels of heterogeneity that we explore, namely heterogeneity across parallel paths and heterogeneity within a single path in the direction of transport. 

Interestingly, the membrane samples studied by Culp et al.~had pathways through the membrane with significantly more heterogeneity \textit{along} pathways than \textit{between} pathways. All pathways within a membrane sample were similar with highly correlated flux. However, each of the parallel pathways had regions of low and high local flux, corresponding to high and low barriers, respectively. Sections with low local flux (high outlier barriers) significantly limited the total flux along all paths. On the other hand, the high permeability membranes had narrower distributions of local flux along the transport coordinate. Consequently, the most permeable membranes had barrier distributions with low variance, such that there were not many high outliers. All paths were similar, and the highest barriers in these paths were relatively low. The correlations in paths and barriers in the work of Culp et al.~differed from those explored in this study. These differences in heterogeneity are likely due to the choices in membrane synthesis or some deeper conceptual reason beyond the scope of this paper. However, we emphasize their observations are still well-described by our overall framework and support our conclusions. 

\section{Conclusions}

In this work, we find that even moderate, statistically random heterogeneity in energy barriers will significantly impact how we interpret the mechanisms of transport through the membrane. In polymeric membranes used for RO and NF, structural and chemical heterogeneity, such as non-uniform voids or charged functional groups, introduce a wide variety of free energy barriers to permeability~\cite{freger_nanoscale_2003,shefer_enthalpic_2021}. We found that for a cross-section of membrane, the overall effective barrier is most dependent on the highest barriers in the most permeable paths with smaller contributions from the other parallel paths. The enthalpic and entropic components, and thus the overall free energy barrier, increase with increasing heterogeneity in the membrane.

Our results suggest that to design membranes with desired separation capabilities we must control the highest barriers to transport in the most permeable paths. Even molecular-level defects or voids in the membrane along the transport coordinate will significantly increase the permeability by decreasing the highest barriers to transport through individual paths, leading to flux hot spots~\cite{ramon_transport_2013}. Designing membranes with more uniform energy barriers, even at the nanoscale level, will distribute flow through more paths. Such nanoscale homogeneity could potentially be achieved through processes such as self-assembly of monomers into nanochannels or regulation of polymerization~\cite{coscia_understanding_2019, zhao_anhydrous_2023, shen_polyamide-based_2022}  Experimentally, increasing the homogeneity of the polyamide films has been shown to increase water flux and permselectivity~\cite{hailemariam_reverse_2020, shen_polyamide-based_2022}. 

There are a number of limitations to this study due to the approximations required to perform numerical experiments, but the framework is flexible enough that the main conclusions are broadly applicable. Some extensions to the theory are also possible; in this framework, we do not consider the coupled diffusion of multiple species, which can constrain transport via requirements of electroneutrality for ions, but recent work has applied the transition-theory framework to coupled multicomponent fluxes~\cite{kingsbury_kinetic_2024}. We also do not include external driving forces in our analysis, but previous work has developed this theory, which effectively scales the energy barriers~\cite{zwolinski_diffusion_1949,del_castillo_energy-barrier_1979}. The magnitudes of the barriers would change, but our interpretation of effective energy barriers would not. 

To better understand the underlying distribution of barriers in polymeric membranes, it is necessary to correlate nanoscale transport phenomena to measured effective barriers. For example, Culp et al.~\cite{culp_nanoscale_2021} were able to quantify the nanoscale heterogeneity in RO membranes, and relating these polymer density  distributions to the effective barriers to transport would provide a sense of scale for the variances relevant in RO and NF transport. Molecular simulations can give examples of molecular mechanisms, but it is necessary to ensure these simulations are representative of physical membrane systems. 

Our work shows that the conventional theoretical framework for transition-state theory energy barriers leads to incorrect interpretations of experimental effective free energy, enthalpic, and entropic barriers. That is, even analysis of experimental results based on transition-state theory will produce effective barriers that are not easily related to mechanistic details at the atomistic level. In particular, the observed effective enthalpies and entropies do not necessarily correspond to either the most frequent or the highest mechanistic barriers occurring in the system. For example, attempts to match barriers to specific enthalpies of ion dehydration within the membrane are unlikely to be successful, as the free energy barrier of an individual mechanistic event may be several kcal/mol different from the measured effective free energy barrier, and thus chemical design attempts may focus on the wrong interactions. Additionally, attempts to understand membrane barriers by looking at typical events in the membrane via simulation may focus on the wrong events, as the typical free energy barrier is not necessarily relevant in the overall permeability. Similarly, the highest barriers encountered within the membrane may not be relevant, as it is only the highest barriers on the most permeable paths that primarily contribute to the experimentally observable barrier. 

\section*{CRediT authorship contribution statement}
\noindent \textbf{Nathanael S. Schwindt:} Methodology, Software, Investigation, Writing - Original Draft, Writing - Review \& Editing \textbf{Mor Avidar:} Investigation \textbf{Razi Epsztein:} Conceptualization, Methodology, Resources, Writing - Review \& Editing, Supervision, Funding acquisition \textbf{Anthony P. Straub:} Conceptualization, Writing - Review \& Editing, Funding acquisition \textbf{Michael R. Shirts:} Conceptualization, Resources, Writing - Review \& Editing, Supervision, Funding acquisition

\section*{Declaration of Competing Interest}
\noindent The authors declare that they have no known competing financial interests or personal relationships that could have appeared to influence the work reported in this paper.

\section*{Acknowledgments}
\noindent Funding: This material is based upon work supported by the National Science Foundation under Grant No. CBET-2136835 and the United States-Israel Binational Science Foundation (BSF), Jerusalem, Israel (grant No. 2021615).


\bibliographystyle{elsarticle-num}
\bibliography{eyring_model}

\clearpage
\pagebreak

\captionsetup[figure]{labelfont={bf}, labelsep=period, name={Fig.}}
\captionsetup[table]{labelfont={bf}, labelsep=period, name={Table}}
\renewcommand\thesection{S\arabic{section}}
\renewcommand\thefigure{S\arabic{figure}}
\renewcommand\thetable{S\arabic{table}}
\renewcommand\theequation{S\arabic{equation}}

\setcounter{figure}{0}
\setcounter{equation}{0}
\setcounter{page}{1}
\setcounter{section}{0}

\section*{\Large Supplementary Materials}

\section{Additional Derivations}

\subsection{Previous theoretical framework} \label{s:SM_previous_framework}
The original derivation by Zwolinski and coworkers~\cite{zwolinski_diffusion_1949} modeled membrane flux in terms of point-to-point jumps of molecules governed by rate constants. Thus, the net flux ($Q$) between equilibrium positions within the membrane becomes the difference in the forward ($k$) and backward ($k'$) molecular jump rates through a cross-sectional area. A single barrier with equal forward and backward rate constants ($k$) and jump lengths ($\lambda$) leads to Fick's first law of diffusion (Eq.~\ref{eq:SM_ficks_law}) with diffusion coefficient $D = k\lambda^2$.

\begin{equation}
    Q = -D \frac{\text{d}C}{\text{d}x}
    \label{eq:SM_ficks_law}
\end{equation}

At steady state, the flux is a set of rate equations relating all local equilibrium positions along the direction of transport. Assuming a constant flux across the membrane and eliminating all the local concentrations gives an expression for the flux in terms of the local rate constants $k_i$, jump lengths $\lami{i}$, and initial $C_0$ and final $C_{n+1}$ concentrations shown in Eq.~\ref{eq:SM_flux1}, where $n$ is the total number of jumps along the transport coordinate.

\begin{equation}
    Q = \frac{ \displaystyle k_0 \lami{0} C_0 - \prod_{i=1}^n \parfrac{k_i' \lami{i}'}{k_i \lami{i}} k_{n+1}' \lami{n+1}' C_{n+1} } { \displaystyle 1 +  \sum_{r=1}^n \prod_{i=1}^r \parfrac{k_i' \lami{i}'}{k_i \lami{i}} }
    \label{eq:SM_flux1}
\end{equation}

\noindent Under transition state theory, the individual rate constants $k_i$ can be related to free energy barriers $\Delta G^{\ddagger}_i$ by 

\begin{equation}
    k_i = \kappa_i \frac{k_B T}{h} \exp{\parfrac{-\Delta G^{\ddagger}_i}{R T}}
    \label{eq:SM_TST_rate_constant1}
\end{equation}

\noindent $\kappa_i$ is the transmission coefficient (generally assumed to be unity for membrane processes), and $k_B$, $T$, and $h$ are Boltzmann's constant, temperature, and Planck's constant, respectively. Zwolinski et al.~\cite{zwolinski_diffusion_1949} and later del Castillo et al.~\cite{del_castillo_energy-barrier_1979} expanded the expression for flux in terms of free energy barriers to include external forces. Here, we explore the model without external forces, as external forces will only increase or decrease the free energy barriers without impacting the behavior of the model. 

Zwolinksi and coworkers verified their model on biological membranes using a simple setup with four distinct rate constants for the solution $k_s$, the solution-membrane interface $k_{sm}$, the membrane $k_m$, and the membrane-solution interface $k_{ms}$. The authors evaluated Eq.~\ref{eq:SM_flux1} for the solution-membrane-solution scenario under the assumptions that all jump lengths are equal, all free energy barriers within the membrane are equal, and diffusion within the membrane is the dominating step. They arrived at the following equation for membrane permeability ($P$)

\begin{equation}
    P = \frac{k_{sm} k_m \lambda}{M k_{ms}}
\end{equation}

\noindent $M$ is the number of membrane jumps. As a result, they expressed membrane permeability in terms of a single, effective free energy barrier that includes the solution-membrane, membrane, and membrane-solution barriers. They claimed that this effective free energy barrier represents the difference in free energy between the species in solution and the species at the top of the highest potential energy barrier within the membrane. They extracted the enthalpic ($\Delta H_{eff}^{\ddagger}$) and entropic ($\Delta S_{eff}^{\ddagger}$) contributions to permeability from the Gibbs-Helmholtz relation. 

\begin{eqnarray}
    P &=& \parfrac{\lambda^2}{\delta} \parfrac{k_B T}{h} \exp{\parfrac{-\Delta G_{eff}^{\ddagger}}{R T}} \nonumber \\ 
    &=& \parfrac{\lambda^2}{\delta} \parfrac{k_B T}{h} \exp{\parfrac{\Delta S_{eff}^{\ddagger}}{R}} \exp{\parfrac{-\Delta H_{eff}^{\ddagger}}{R T}}
    \label{eq:SM_P_eyring1949}
\end{eqnarray}

$\delta$ in Eq.~\ref{eq:SM_P_eyring1949} is the membrane thickness, defined as $\delta = M\lambda$. This expression has been applied to both biological and polymeric membrane systems as a way to explore the molecular mechanisms governing membrane permeability~\cite{lopez_enthalpic_2017,shefer_enthalpic_2021,shefer_applying_2022}.

Giddings and Eyring also explored barrier kinetics primarily through the lens of nucleation~\cite{giddings_multi-barrier_1958}. Starting from Eq.~\ref{eq:SM_flux1}, the authors represented the effective free energy barrier for flux in terms of the individual point-to-point rate constants. While they did not explicitly state the similarity, the effective free energy barrier is in the form of multiple parallel resistances (see Equation 7 in reference~\cite{giddings_multi-barrier_1958}). They developed a “kT-cutoff model” to identify the non-negligible barriers (i.e. those within $k_B T$ of the maximum barrier). They concluded that for a series of jumps over unequal free energy barriers, the highest barrier does not define the overall flux but rather contributes the most to a sum of non-negligible barriers. Furthermore, they showed that the effective barrier depends only on the magnitude of the contributing barriers, not on their order. 

Scheuplein further explored the idea that position does not matter in his analysis of Gidding and Eyring’s multibarrier kinetics model more specifically applied to membrane permeability~\cite{scheuplein_application_1968}. Scheuplein grouped membrane barriers of similar size and represented membrane transport across many unequal groups as transport across a series of membranes with equal barriers. For barrier groups $\alpha, \beta, ..., \omega$, the permeability becomes 

\begin{equation}
    \frac{1}{P} = \parfrac{M}{\lambda} \sum_{i=\alpha}^{\omega} \parfrac{p_i}{K_{si} k_i}
\end{equation}

\noindent $M$ is the total number of barriers, $K_{si}$ is the partition coefficient from the solution to the $i^{th}$ minimum in the membrane, and $p_i$ is the probability of occurrence of the $i^{th}$ kind of barrier. This representation shows that the permeability is dependent on the individual probabilities and rate constants within the membrane. Therefore, the permeability is most affected by the highest and the most probable membrane barriers. 

This equation leads to interpreting membrane permeability as combining parallel resistances. Wendt et al.~derived a similar interpretation of permeability for pores in series~\cite{wendt_effect_1976}, and del Castillo et al.~explicitly showed how the multibarrier kinetic model can be thought of under this context~\cite{del_castillo_energy-barrier_1979}. Wendt and coworkers' primary assumptions were that transport can be treated as one-dimensional and that there is no internal concentration polarization within the membrane. From these assumptions, they showed that the overall flux in a series of non-sieving pores is equivalent to flux through a single pore with an overall permeability in the form of parallel resistances. Expanding to an array of pores, they showed that the overall flux in parallel pores is a sum of the individual pore fluxes. The overall permeability for the parallel array of pores is the area-weighted sum of the $n$ individual pore permeabilities ($P_i$) as shown in Eq.~\ref{eq:SM_parallel_permeability}. 
\begin{equation}
    P = \sum_{i=1}^N \frac{A_i}{A_0} P_i
    \label{eq:SM_parallel_permeability}
\end{equation}

\noindent where $A_i$ is the individual pore area and $A_0$ is the total membrane area considered, generally assumed to be a unit area. The individual pore areas are not required to sum to the total area. As a result, the overall permeability $P$ describes transport through the accessible area. If the pore areas do sum to the total area, the overall permeability becomes a weighted average, and the entire  membrane area is accessible for transport. del Castillo et al.~also explored these permeability expressions under arbitrary external forces, arguing that the overall flux depends on the distribution of parallel permeabilities, but in most cases, it will be near the pure diffusion limit. Additionally, they provided a weak constraint on the applicability of the multibarrier kinetic model for membrane transport. 

\subsection{Derivation of the permeability with distributions of barriers, jumps, and paths} \label{s:SM_permeability_derivation}
To construct our framework, we start with the main assumptions of Eyring's multibarrier kinetic model applied to the solution-membrane-solution scenario, and then relax some of these assumptions. Their assumptions were: 
\begin{enumerate}
    \item steady state flux can be represented by point-to-point molecular jumps between locally equilibrated states,
    \item membrane transport is one-dimensional,
    \item all solution jumps have equal rate constants and jump lengths,
    \item an aqueous solution is diffusing through the membrane, and membrane diffusion is the primary hindrance to transport,
    \item the transmission coefficient is one for all rate constants,
    \item the free energy barriers within the membrane are a series of equal free energy barriers, and 
    \item the jump lengths between local barriers are equal.
\end{enumerate}

We start with Eq.~\ref{eq:SM_flux2} in the same way as Zwolinski et al.~\cite{zwolinski_diffusion_1949}, but we do not apply the assumptions that the free energy barriers within the membrane are a series of equal free energy barriers and the jump lengths between local barriers are equal.

\begin{equation}
    Q = \frac{ \displaystyle k_0 \lami{0} C_0 - \prod_{i=1}^n \parfrac{k_i' \lami{i}'}{k_i \lami{i}} k_{n+1}' \lami{n+1}' C_{n+1} } { \displaystyle 1 +  \sum_{r=1}^n \prod_{i=1}^r \parfrac{k_i' \lami{i}'}{k_i \lami{i}} }
    \label{eq:SM_flux2}
\end{equation}

\noindent For the solution-membrane-solution scenario, we define four kinds of jumps. We use a solution jump with rate constant $k_s$ and jump length $\lambda_{s}$, a solution-membrane interfacial jump with rate constant $k_{sm}$ and jump length $\lambda_{sm}$, a series of membrane jumps with rate constants $k_{m,i}$ and jump lengths $\lambda_{m,i}$, and a membrane-solution interfacial jump with rate constant $k_{ms}$ and jump length $\lambda_{ms}$. As a result, the numerator expands to 

\begin{align*}
     & k_0 \lami{0} C_0 - \prod_{i=1}^n \parfrac{k_i' \lami{i}'}{k_i \lami{i}} k_{n+1}' \lami{n+1}' \lambda C_{n+1} \\
     &= k_s \lami{s} C_0 - \left[ \parfrac{k_s \lami{s}}{k_s \lami{s}} ... \parfrac{k_s \lami{s}}{k_s \lami{s}} \parfrac{k_s \lami{s}}{k_{sm} \lami{sm}} \parfrac{k_{ms} \lami{ms}}{k_{m,1} \lami{m,1}} \parfrac{k_{m,1} \lami{m,1}}{k_{m,2} \lami{m,2}} ... \right. \\
     & \hspace{60px} \left. \parfrac{k_{m,M-1} \lami{m,M-1}}{k_{m,M} \lami{m,M}} \parfrac{k_{m,M} \lami{m,M}}{k_{ms} \lami{ms}} \parfrac{k_{sm} \lami{sm}}{k_s \lami{s}} \parfrac{k_s \lami{s}}{k_s \lami{s}} ... \parfrac{k_s \lami{s}}{k_s \lami{s}} \right] k_s \lami{s} C_{n+1} \\
     &=  k_s \lami{s} C_0 - \left[ \parfrac{k_s \lami{s}}{k_s \lami{s}}^{S-1} \parfrac{k_s \lami{s}}{k_{sm} \lami{sm}} \parfrac{k_{ms} \lami{ms}}{k_{m,1} \lami{m,1}} \parfrac{k_{m,1} \lami{m,1}}{k_{m,2} \lami{m,2}} ... \right. \\
     & \hspace{60px} \left. \parfrac{k_{m,M-1} \lami{m,M-1}}{k_{m,M} \lami{m,M}} \parfrac{k_{m,M} \lami{m,M}}{k_{ms} \lami{ms}} \parfrac{k_{sm} \lami{sm}}{k_s \lami{s}} \parfrac{k_s \lami{s}}{k_s \lami{s}}^{S'-1} \right] k_s \lami{s} C_{n+1} \\
     &= k_s \lami{s} C_0 - \left( 1 \right) k_s \lami{s} C_{n+1}
\end{align*}

\noindent Here, there are $S$ solution jumps before the membrane, $M$ membrane jumps, $S'$ solution jumps after the membrane, and $n$ total jumps. 

\pagebreak

\noindent The denominator expands to 

\begin{align*}
    & 1 + \sum_{r=1}^n \prod_{i=1}^r \parfrac{k_i' \lami{i}'}{k_i \lami{i}} \\
    &= 1 + \left( S-1 \right) \left[ \parfrac{k_s \lami{s}}{k_s \lami{s}}^{S-1} \right]_{S-1} + \left[ \parfrac{k_s \lami{s}}{k_s \lami{s}}^{S-1} \parfrac{k_s \lami{s}}{k_{sm \lami{sm}}} \right]_{S} \\
    & \hspace{20px} + \left[ \parfrac{k_s \lami{s}}{k_s \lami{s}}^S \parfrac{k_s \lami{s}}{k_{sm } \lami{sm}} \parfrac{k_{ms} \lami{ms}}{k_{m,1} \lami{m,1}} \right]_{S+1} \\
    & \hspace{20px} + \left[ \parfrac{k_s \lami{s}}{k_s \lami{s}}^{S-1} \parfrac{k_s \lami{s}}{k_{sm} \lami{sm}} \parfrac{k_{ms} \lami{ms}}{k_{m,1} \lami{m,1}} \parfrac{k_{m,1} \lami{m,1}}{k_{m,2} \lami{m,2}} \right]_{S+2} \\ 
    & \hspace{20px} + ... + \left[ \parfrac{k_s \lami{s}}{k_s \lami{s}}^{S-1} \parfrac{k_s \lami{s}}{k_{sm} \lami{sm}} \parfrac{k_{ms} \lami{ms}}{k_{m,1} \lami{m,1}} \parfrac{k_{m,1} \lami{m,1}}{k_{m,2} \lami{m,2}} ... \right. \\
    & \hspace{65px} \left. \parfrac{k_{m,M-1} \lami{m,M-1}}{k_{m,M} \lami{m,M}} \parfrac{k_{m,M} \lami{m,M}}{k_{ms} \lami{ms}} \right]_{S+M+1} \\
    & \hspace{20px} + \left[ \parfrac{k_s \lami{s}}{k_s \lami{s}}^{S-1} \parfrac{k_s \lami{s}}{k_{sm} \lami{sm}} \parfrac{k_{ms} \lami{ms}}{k_{m,1} \lami{m,1}} \parfrac{k_{m,1} \lami{m,1}}{k_{m,2} \lami{m,2}} ... \right. \\
    & \hspace{40px} \left. \parfrac{k_{m,M-1} \lami{m,M-1}}{k_{m,M} \lami{m,M}} \parfrac{k_{m,M} \lami{m,M}}{k_{ms} \lami{ms}} \parfrac{k_{sm} \lami{sm}}{k_s \lami{s}} \right]_{S+M+2} \\
    & \hspace{20px} + \left( S'-1 \right) \left[ \parfrac{k_s \lami{s}}{k_s \lami{s}}^{S-1} \parfrac{k_s \lami{s}}{k_{sm} \lami{sm}} \parfrac{k_{ms} \lami{ms}}{k_{m,1} \lami{m,1}} \parfrac{k_{m,1} \lami{m,1}}{k_{m,2} \lami{m,2}} ... \right. \\
    & \hspace{80px} \left. \parfrac{k_{m,M-1} \lami{m,M-1}}{k_{m,M} \lami{m,M}} \parfrac{k_{m,M} \lami{m,M}}{k_{ms} \lami{ms}} \parfrac{k_{sm} \lami{sm}}{k_s \lami{s}} \parfrac{k_s \lami{s}}{k_s \lami{s}}^{S'-1} \right]_{S+M+S'+1} \\
    &= 1 + \left( S-1 \right) + \parfrac{k_s \lami{s}}{k_{sm} \lami{sm}} + \parfrac{k_s \lami{s} k_{ms} \lami{ms}}{k_{sm} \lami{sm} k_{m,1} \lami{m,1}} + \parfrac{k_s \lami{s} k_{ms} \lami{ms}}{k_{sm} \lami{sm} k_{m,2} \lami{m,2}} + ... \\
    & \hspace{20px} + \parfrac{k_s \lami{s} k_{ms} \lami{ms}}{k_{sm} \lami{sm} k_{m,M} \lami{m,M}} + \parfrac{k_s \lami{s}}{k_{sm} \lami{sm}} + 1 + \left( S'-1 \right) \\
    &= S + S' + 2 \parfrac{k_s \lami{s}}{k_{sm} \lami{sm}} + \parfrac{k_s \lami{s} k_{ms} \lami{ms}}{k_{sm} \lami{sm}} \sum_{j=1}^M \parfrac{1}{k_{m,j} \lami{m,j}}
\end{align*}

\noindent where the subscripts on bracketed terms track the sum over all jumps. Eq.~\ref{eq:SM_flux2} simplifies to

\begin{equation}
    Q = \frac{k_s \lami{s} \left( C_0 - C_{n+1} \right)}{\displaystyle S + S' + 2 \parfrac{k_s \lami{s}}{k_{sm} \lami{sm}} + \parfrac{k_s \lami{s} k_{ms} \lami{ms}}{k_{sm} \lami{sm}} \sum_{j=1}^M \parfrac{1}{k_{m,j} \lami{m,j}}}
    \label{eq:SM_flux_final}
\end{equation}

\noindent Therefore, the permeability as defined in the original derivation by Zwolinski et al.~becomes

\begin{equation}
    P = \frac{1}{\displaystyle \parfrac{S}{k_s \lami{s}} + \parfrac{S'}{k_s \lami{s}} + \parfrac{2}{k_{sm} \lami{sm}} + \parfrac{k_{ms} \lami{ms}}{k_{sm} \lami{sm}} \sum_{j=1}^M \parfrac{1}{k_{m,j} \lami{m,j}}}
\end{equation}

\noindent In polymeric membrane transport, the jump rates through solution ($k_s$) are significantly larger than those through the membrane interface and the bulk membrane, since motion in the membrane is significantly hindered compared to motion in solution~\cite{zwolinski_diffusion_1949}. As a result, the permeability can be expressed only in terms of the interfacial and membrane rate constants as shown in Eq.~\ref{eq:SM_permeability_rates}.

\begin{equation}
    \frac{1}{P} = \parfrac{2}{k_{sm} \lami{sm}} + \parfrac{k_{ms} \lami{ms}}{k_{sm} \lami{sm}} \sum_{j=1}^M \parfrac{1}{k_{m,j} \lambda_{m,j}}
    \label{eq:SM_permeability_rates}
\end{equation}

\noindent The first term in Eq.~\ref{eq:SM_permeability_rates} is associated with diffusion through the solution-membrane interface, and the second term is associated with diffusion through the membrane. For most polymeric membranes, the rate-determining step is diffusion through the membrane~\cite{zhou_intrapore_2020}, so Eq.~\ref{eq:SM_permeability_rates} can be approximated with only the second term. The resulting expression for permeability in terms of the rate constants for transport is shown in Eq.~\ref{eq:SM_membrane_diffusion_limited_permeability}.

\begin{equation}
    P = \frac{k_{sm} \lami{sm}}{\displaystyle k_{ms} \lami{ms} \sum_{j=1}^M \parfrac{1}{k_{m,j} \lambda_{m,j}}}
    \label{eq:SM_membrane_diffusion_limited_permeability}
\end{equation}

\noindent Under transition state theory, the individual rate constants $k_i$ can be related to free energy barriers $\Delta G^{\ddagger}_i$ by 

\begin{equation}
    k_i = \kappa_i \frac{k_B T}{h} \exp{\parfrac{-\Delta G^{\ddagger}_i}{k_B T}}
    \label{eq:SM_TST_rate_constant2}
\end{equation}

\noindent Relating Eq.~\ref{eq:SM_membrane_diffusion_limited_permeability} to the associated free energy barriers with Eq.~\ref{eq:SM_TST_rate_constant2} yields Eq.~\ref{eq:SM_permeability_barriers} for permeability across a series of unequal membrane barriers in terms of the free energy barriers at transition $\Delta G_{m,j}^{\ddagger}$.

\begin{equation}
    P = \frac{\displaystyle \parfrac{\lambda_{sm}}{\lambda_{ms}} \parfrac{k_B T}{h} \exp{\parfrac{ -\left( \Delta G_{sm}^{\ddagger} - \Delta G_{ms}^{\ddagger} \right)}{R T}}}{\displaystyle \sum_{j=1}^M \parfrac{1}{\lambda_{m,j} \exp{\parfrac{-\Delta G_{m,j}^{\ddagger}}{R T}}}}
    \label{eq:SM_permeability_barriers}
\end{equation} 

\subsection{Derivation of the effective free energy barrier}
Zwolinski and coworkers express the effective free energy barrier to permeability as 

\begin{eqnarray}
    P &=& \parfrac{\lambda^2}{\delta} \parfrac{k_B T}{h} \exp{\parfrac{-\Delta G_{eff}^{\ddagger}}{R T}} \nonumber \\ 
    &=& \parfrac{\lambda^2}{\delta} \parfrac{k_B T}{h} \exp{\parfrac{\Delta S_{eff}^{\ddagger}}{R}} \exp{\parfrac{-\Delta H_{eff}^{\ddagger}}{R T}}
    \label{eq:SM_permeability_eyring1949}
\end{eqnarray}

\noindent We incorporate parallel molecular pathways and distributions of membrane jumps and barriers into the transition-state theory model for membrane permeability by applying the single path permeability in Eq.~\ref{eq:SM_permeability_barriers} to the overall permeability for a parallel array of paths in Eq.~\ref{eq:SM_parallel_permeability}. The resulting equation for overall permeability across $N$ parallel paths is shown in Eq.~\ref{eq:SM_overall_permeability}.

\begin{equation}
    P = \sum_{i=1}^{N} \left[ \frac{\displaystyle \parfrac{A_i}{A_0} \parfrac{\lambda_{sm}}{\lambda_{ms}} \parfrac{k_B T}{h} \exp{\parfrac{ -\left( \Delta G_{sm}^{\ddagger} - \Delta G_{ms}^{\ddagger} \right)}{R T}}}{\displaystyle \sum_{j=1}^{M_i} \parfrac{1}{\lambda_{m,i,j} \exp{\parfrac{-\Delta G_{m,i,j}^{\ddagger}}{R T}}}} \right]
    \label{eq:SM_overall_permeability}
\end{equation}

\noindent $\Delta G_{m,i,j}^{\ddagger}$ and $\lambda_{m,i,j}$ are the free energy barrier and the jump length associated with the $j^{th}$ jump in the $i^{th}$ path. $M_i$ is the number of jumps for path $i$. We equate these expressions for permeability and solve for the effective free energy barrier in terms of the distributions of membrane barriers and parallel paths. 

\begin{equation}
    \Delta G_{eff}^{\ddagger} = -R T \ln{\left[ \sum_{i=1}^{N} \frac{ \displaystyle \parfrac{A_i}{A_0} \parfrac{\delta}{\lami{avg}^2} \parfrac{\lambda_{sm}}{\lambda_{ms}} }{\displaystyle \sum_{j=1}^{M_i} \parfrac{1}{\lambda_{m,i,j}} \exp{\parfrac{\Delta G_{m,i,j}^{\ddagger}}{R T}}} \right]} + \left( \Delta G_{sm}^{\ddagger} - \Delta G_{ms}^{\ddagger}\right)
    \label{eq:SM_effective_barrier}
\end{equation} 

\section{Estimation and Sensitivity of Model Parameters}

\subsection{Estimating the number of paths per unit area} \label{s:estimate_paths}
We estimate the number of paths per unit area for the polyamide membrane to be an order of magnitude fewer than what is expected for single-walled carbon nanotubes (SWCNT). SWCNT with diameter 1.7~nm have been reported to pack with density \num{1.9e12} paths per \unit{\cm\squared}~\cite{jue_ultrapermeable_2020}. If all the area is occupied by circular nanotubes with diameter 1.7~nm and negligible thickness, the theoretical packing density is \num{4.4e13} paths per \unit{\cm\squared}. We use this ratio of actual packing density to theoretical packing density to approximate the actual packing density of SWCNT with diameter 0.5~nm, the reported average pore size for polyamide membranes~\cite{chu_variation_2021, wang_water_2023}. 

\begin{equation}
    \left( \frac{\text{actual, } d=1.7 \unit{\nm}}{\text{theoretical, } d=1.7 \unit{\nm}} \right) \times \left( \text{theoretical, } d=0.5 \unit{\nm} \right) = \left( \text{actual, } d=0.5 \unit{\nm} \right)
\end{equation}

\noindent We estimate the actual packing density of SWCNT with diameter 0.5~nm to be \num{2.2e13} paths per \unit{\cm\squared} or 0.22 paths per \unit{\nm\squared}. Therefore, we estimate the actual number of paths per unit area for polyamide membranes to be 0.022 paths per \unit{\nm\squared}. The results we present consider a total unit area of 0.1 \unit{\um\squared}, or \num{1.0e7} $\text{\AA}^2$, and a single path area of $\pi (5 \text{\AA})^2 = 19.635~\text{\AA}^2$. These areas correspond to 2196 paths per 0.1 \unit{\um\squared}. 

\begin{figure}[H]
    \centering
    \includegraphics[width=\textwidth]{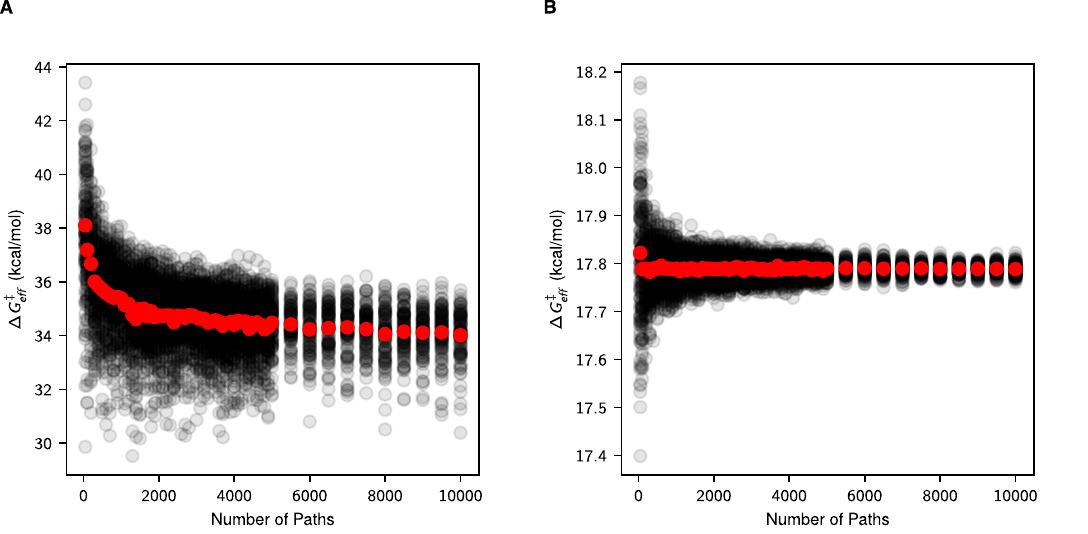}
    \caption{\textbf{{Convergence testing to determine necessary number of paths.}} The effective free energy barrier converges for both normally (\textbf{A}) and exponentially (\textbf{B}) distributed barriers for the distribution variances used in this paper converges within summation over 2000 pathways. Barrier distributions with higher variance will take more pathways to converge. Black points are single realizations of calculated effective free energy barriers, and red points are effective free energy barriers averaged over all realizations. We calculate 300 realizations for each number of paths.}
    \label{fig:convergence}
\end{figure}

\subsection{Effect of jump distributions on the effective free energy barrier} \label{s:jump_distributions}
Given a fixed membrane thickness, the distribution of number of jumps and the length of jumps are directly related. Thus, we can examine the effects of only the distribution in the number of jumps for a given membrane thickness. We choose a physically realistic membrane thickness and hold all membrane free energy barriers equal. We draw the number of jumps from a (truncated) normal distribution because we can change the variance while maintaining a physically relevant mean. For this analysis, we model 5000 paths through a membrane of thickness 400 \AA. We set the mean number of jumps to be 100, and we adjust the jump length to ensure the membrane thickness remains constant. We vary the standard deviation in the number of jumps between 5 and 200. Because this can result in a negative number of jumps, we redraw each negative draw from a normal distribution until no paths have a non-positive number of jumps. This results in a nearly normal distribution for large variances, but a truncated distribution at $N=1$ and below for larger variances. 

\begin{figure}[ht!]
    \centering
    \includegraphics[width=\textwidth]{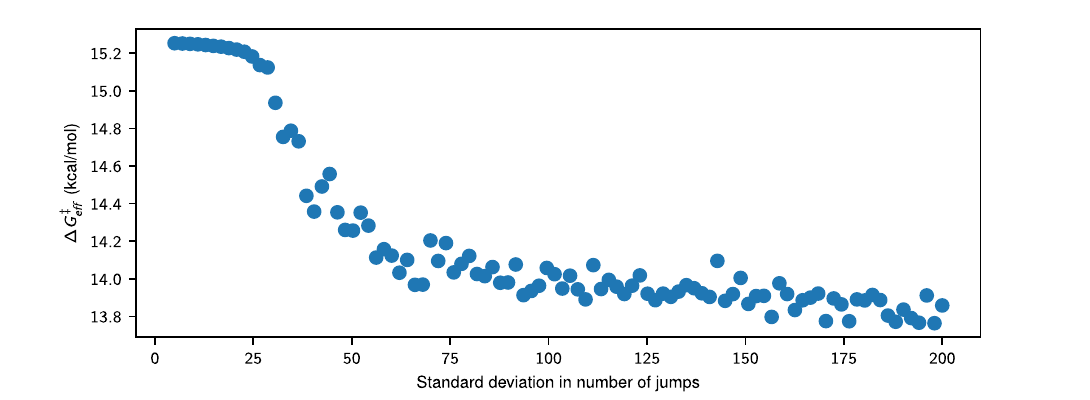}
    \caption{\textbf{Effective barrier decreases with increasing variance in the number of jumps.} We show the overall effective free energy barrier as a function of the standard deviation for normally distributed numbers of jumps with mean 100. The jump length is adjusted to maintain a constant membrane thickness of 40 \AA. For each standard deviation, we calculate the overall effective free energy barrier over 5000 paths.}
    \label{fig:jump_variances}
\end{figure}

We find that the effective free energy barrier decreases with increasing variance in the number of jumps, but the change is significantly less than the effects from distributions of barrier heights in all physical scenarios. Fig.~\ref{fig:jump_variances} shows the relationship between the effective free energy barrier and the standard deviation in number of jumps. 

The effective barrier decreases negligibly when the variance is small. When the variance becomes larger, a small number of paths have very few jumps, which results in moderately decreased effective barrier, up to 1.5 kcal/mol. The barrier decreases negligibly again for larger standard deviations of the truncated distribution, when the number of paths with $N=1$ barriers predominates. In contrast, modest variance in the barrier height distribution, as shown by the normally distributed barrier heights in Fig.~3B in the main text, changes the effective barrier by 5.3 kcal/mol.

\begin{figure}[ht!]
    \centering
    \includegraphics[width=\textwidth]{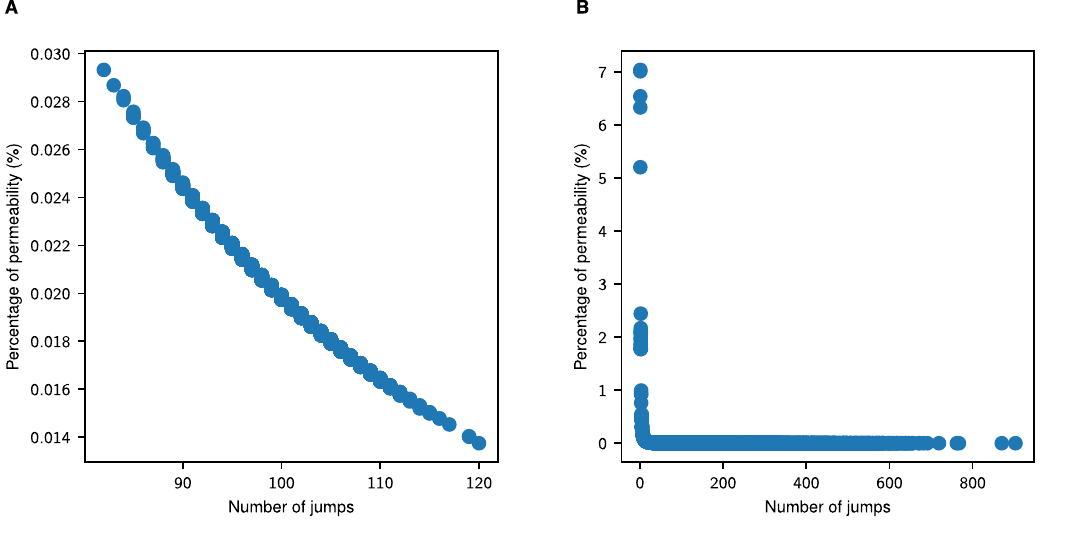}
    \caption{\textbf{Paths with only a few jumps contribute most to the permeability.} We calculate the percentage of the total permeability through paths with different numbers of jumps. The jump length is adjusted to maintain a constant membrane thickness of 40 \AA, and we calculate the overall effective free energy barrier over 5000 paths. (\textbf{A}) The standard deviation in the normally distributed number of jumps is 5, and the mean is 100. The permeability is evenly spread across paths, since all paths have similar number of jumps. (\textbf{B}) The standard deviation in the normally distributed number of jumps is 200, and the mean is 100. The permeability is dominated by paths with only a few jumps.}
    \label{fig:perm_percentage}
\end{figure}

At large standard deviation, the effective barrier becomes dominated by paths with only a few jumps. Fig.~\ref{fig:perm_percentage}A confirms this trend by showing the percentage of the total permeability through each path for the highest variance distribution (standard deviation 200). Conversely, when all paths have nearly the same number of jumps, the permeability is evenly distributed across the paths, as shown in Fig.~\ref{fig:perm_percentage}B. The standard deviation in the number of jumps for Fig.~\ref{fig:perm_percentage}B is 5. 

Physically, larger jumps along the transport coordinate with the approximately the same membrane thickness reduce the number of barriers the molecules must cross. Jump lengths affect the single path permeability as a sum of reciprocal jump lengths, so small jumps contribute more than large jumps. The distributions of jump lengths introduce some smaller jumps that drive the permeability lower and the effective free energy barrier higher. Individual jump lengths are likely to be correlated with their associated free energy barrier. However, the exponential contribution of the free energy barriers will dominate the contribution from the jump lengths. For membranes with heterogeneity in their free energy barrier distributions, the variability of the smallest maximum barrier contributes significantly more than variability in the number and length of jumps through the membrane, and we thus focus primarily on the distribution of barrier heights in this study. 

\section{Fitting Experimental Data}

\subsection{Accounting for concentration polarization in the membrane} \label{s:concentration_polarization}
Previously reported barriers for NF and RO membranes range from 0 to $\sim$17 kcal/mol with most values lie between $\sim$4 and $\sim$8 kcal/mol~\cite{epsztein_towards_2020}. However, most of the reported values in the literature are likely an overestimation of the real barriers, as these values were measured without accounting for the increasing concentration polarization of the transported solutes with temperature. This phenomenon leads to higher concentration gradient over the membrane (and therefore higher driving force) with temperature, resulting in an increased solute flux that is not related to intrinsic activation (i.e., a permeability increase with temperature). Our measurements rigorously accounted for concentration polarization and therefore reflect more reliably the intrinsic barriers. 

In brief, to account for concentration polarization during the measurement of the permeability at the different temperatures, evaluation of the salt concentration on the membrane surface, $C_m$, was performed at each temperature by retrieving the mass transfer coefficient in the boundary layer, $k$, using the following correlation for the Sherwood number based on laminar (Eq.~\ref{eq:sherwood_laminar}) and turbulent (Eq.~\ref{eq:sherwood_turbulent}) flows in a rectangular channel without a spacer~\cite{mulder_basic_1996}: 
\begin{equation}
    \text{Sh} = 1.85 \left( \text{Re} \text{Sc} \frac{d_h}{L} \right)^{0.33}
    \label{eq:sherwood_laminar}
\end{equation}
\begin{equation}
    \text{Sh} = 0.04 \text{Re}^{0.75} \text{Sc}^{0.33}
    \label{eq:sherwood_turbulent}
\end{equation}
where Sh is the Sherwood number $\left( \text{Sh}=\left( \frac{k d_h}{D} \right) \right)$, Re is the Reynolds number ($\sim$3295 in our system), Sc is the Schmidt number $\left( \text{Sc} = \frac{\nu}{D} \right)$, where $D$ is the diffusion coefficient and $\nu$ is the kinematic viscosity), $d_h$ is the hydraulic radius (\num{1.55e-3}~m in our system), and $L$ is the cell length (0.06~m in our system). The height and width of the flow channel in our system were 0.8~mm and 25~mm, respectively. Because Re was in the borderline of laminar and turbulent flow in our system, we examined both the laminar and turbulent correlations. The diffusion coefficients of the different ions at the tested temperatures were calculated with the Stokes-Einstein equation using Stokes radii (Table~\ref{tab:stokes}). For each salt, the diffusion coefficient of the slower ion was used for the calculations of the Sherwood number. The evaluated $k$ values were then used in the film theory equation (Eq.~\ref{eq:film_theory}) to measure $C_m$.
\begin{equation}
    \frac{C_m - C_p}{C_f - C_p} = \exp{\left( \frac{J_w}{k} \right)}
    \label{eq:film_theory}
\end{equation}
$C_p$ and $C_f$ are the salt concentrations in the permeate and the feed solution, respectively, $J_w$ is the permeate flux (L~m$^{-2}$~h$^{-1}$), and $k$ is the mass transfer coefficient (m~s$^{-1}$).

\begin{table}[ht]
    \centering
    \begin{tabular}{|c|c|}
        \hline
         Species           & Stokes radius (nm)~\cite{nightingale_phenomenological_1959} \\
         \hline
         Sodium (Na$^+$)   & 0.184 \\ 
         \hline
         Fluoride (F$^-$)  & 0.166 \\
         \hline
         Chloride (Cl$^-$) & 0.121 \\
         \hline
    \end{tabular}
    \caption{\textbf{Stokes radii for the ions tested in the experimental filtration measurements.} All data is from reference~\cite{nightingale_phenomenological_1959}}.
    \label{tab:stokes}
\end{table}

\subsection{Comparing the Arrhenius plots and transition-state theory plots}
Energy barriers to permeability are often measured as Arrhenius barriers, and the effective parameters are determined as the slope and intercept of $\ln(P)$ vs $1/T$. However, this form neglects the temperature dependence of the prefactor that is explicitly stated in transition-state theory. The difficulty with the transition-state theory approach is the need for additional parameters, namely average jump length $\lambda$ and membrane thickness $\delta$ in Eq.~\ref{eq:SM_permeability_eyring_effective}, which are challenging to measure. 

\begin{equation}
    P = \parfrac{\lambda^2}{\delta} \parfrac{k_B T}{h} \exp{\parfrac{-\Delta G_{eff}^{\ddagger}}{R T}}
    \label{eq:SM_permeability_eyring_effective}
\end{equation}

\begin{figure}[ht!]
    \centering
    \includegraphics[width=0.75\textwidth]{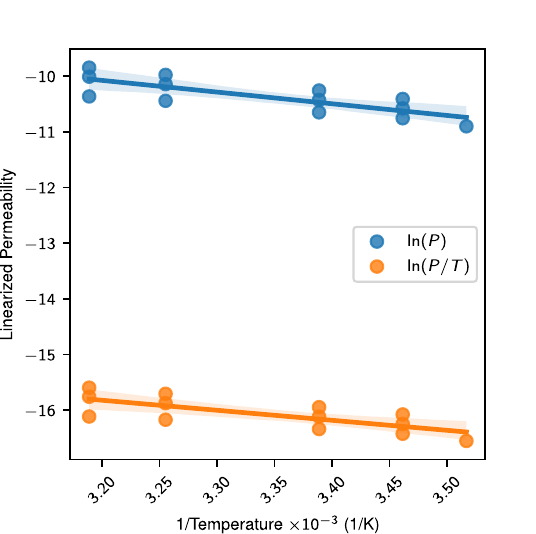}
    \caption{\textbf{Comparison of transition-state theory and Arrhenius plots} Estimation of the effective enthalpic barriers from the transition-state theory model and the Arrhenius model are indistinguishable within error. Scatter points are the experimental data linearized to fit the corresponding model. The least squares fits are shown as lines. A 95\% confidence interval (shaded region) is provided with each least squares fit, determined by a nonparametric bootstrap over 1000 bootstraps.}
    \label{fig:arrhenius_vs_tst}
\end{figure}

\begin{figure}[ht!]
    \centering
    \includegraphics[width=0.75\textwidth]{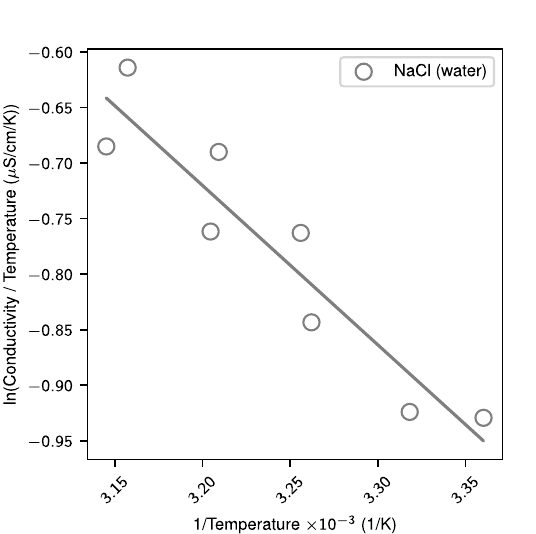}
    \caption{\textbf{Effective enthalpy for NaCl in water.} Linearized transition-state theory plot for the conductivity of NaCl in water, which corresponds to free diffusion of the ions. The least squares fit is shown as a line.}
    \label{fig:tst_NaCl_water}
\end{figure}

We perform both linear fits, and we determine the goodness of fit is not significantly different between the models. The $R^2$ is 0.642 for the Arrhenius treatment and 0.569 for the transition-state theory treatment. In Fig.~\ref{fig:arrhenius_vs_tst}, we show the linear fits for $\ln(P)$ and $\ln(P/T)$ for NaCl in the NF270 membrane. Table~\ref{tab:effective_enthalpies} shows the effective enthalpic barriers for each linearization. The errors shown are the standard errors in the slope parameter for the linear fit propagated to the effective enthalpic barrier. Kinetic theory has shown that the Arrhenius activation energy is related to the TST enthalpic barrier by $\Delta H = E_a - RT$, and our results are consistent with this relationship. Effective enthalpic barriers are the same within error for all systems. The trends in the effective enthalpic barriers and the Arrhenius activation energies are completely preserved. 

\begin{table}
    \centering
    \begin{tabular}{|c|c|c|}
        \hline
        System & Linearization & $\Delta H_{eff}^{\ddagger}$ (kcal/mol) \\
        \hline
        \multirow{2}{*}{NaCl (NF270)} & $\ln (P)$   & $3.6 \pm 0.9$ \\\cline{2-3}
                                      & $\ln (P/T)$ & $3.6 \pm 0.9$ \\
        \hline
        \multirow{2}{*}{NaF (NF270)}  & $\ln (P)$   & $3.9 \pm 1.1$ \\\cline{2-3}
                                      & $\ln (P/T)$ & $3.9 \pm 1.1$ \\
        \hline
        \multirow{2}{*}{NaCl (RO)}    & $\ln (P)$   & $4.4 \pm 1.3$ \\\cline{2-3}
                                      & $\ln (P/T)$ & $4.5 \pm 1.3$ \\
        \hline
    \end{tabular}
    \caption{\textbf{Effective barriers from Arrhenius and transition-state theory models.} The effective enthalpic barrier $\Delta H_{eff}^{\ddagger}$ are the same within error for all systems.}
    \label{tab:effective_enthalpies} 
\end{table}


\end{document}